\documentclass[seceq]{ptptex}

\usepackage{graphicx}

\title{
Dynamics and Accretion of Planetesimals
}

\author{
Eiichiro \textsc{KOKUBO}$^{1}$ and
Shigeru \textsc{IDA}$^{2}$
}

\inst{
$^1$ Division of Theoretical Astronomy, National Astronomical
 Observatory of Japan, \\
 Osawa, Mitaka, Tokyo, 181-8588, Japan \\
$^2$ Department of Earth and Planetary Sciences, Tokyo Institute of
 Technology, \\
Ookayama, Meguro, Tokyo, 152-8551, Japan
}

\abst{
In the standard scenario of planet formation, planets are formed from a
 protoplanetary disk that consists of gas and dust.
The building blocks of solid planets are called as planetesimals that
 are formed by coagulation of dust. 
We review the basic dynamics and accretion of planetesimals by showing
 $N$-body simulations.
The orbits of planetesimals evolve through two-body gravitational
 relaxation: viscous stirring increases the random velocity and
 dynamical friction realizes the equiparation of the random energy.
In the early stage of planetesimal accretion the growth mode of
 planetesimals is runaway growth where larger planetesimals grow faster
 than smaller ones.
When a protoplanet (runaway-growing planetesimal) exceeds a critical
 mass the growth mode shifts to oligarchic growth where similar-sized
 protoplanets grow keeping a certain orbital separation.
The final stage of terrestrial planet formation is collision among
 protoplanets known as giant impacts. 
We also summarize the dynamical effects of disk gas on planets and the
 core accretion model for formation of gas giants and discuss 
 the diversity of planetary systems.
}

\begin{document}

\maketitle

\section{Introduction} 


The solar system consists of planets, their satellites,
 and a huge number of minor bodies (asteroids, trans-Neptunian objects,
 and comets). 
The planets can be classified into three groups: 
 terrestrial planets (Mercury, Venus, Earth and Mars), 
 gas giants (Jupiter and Saturn), and ice giants (Uranus and Neptune).  
These groups differ from one another by compositions, planetary masses,
 and orbital radii. 
The terrestrial planets are light (masses $M \lesssim M_\oplus$, where 
 $M_\oplus$ is the Earth mass, $6 \times 10^{27}$ g), rocky ones with
 relatively small orbital radii ($a \lesssim 1$ AU, where AU is the
 astronomical unit that is the average distance between the sun and the
 Earth, $1.5 \times 10^{13}$ cm), gas giants are heavy 
 ($M \gtrsim 10^2 M_{\oplus}$) planets with main components of H/He gas
 and $a \sim 5$--10 AU, and the ice giants are moderately massive 
 ($M \sim 10M_{\oplus}$ with main components of
 H$_\mathrm{2}$O/CH$_\mathrm{4}$/NH$_\mathrm{3}$ ice and are in distant
 regions ($a \sim 20$--30 AU).  
These planetary orbits are nearly circular and coplanar, which suggests
 that the solar system was formed from a protoplanetary disk around the
 proto-sun.  

The standard scenario for solar system formation was established in
 1960's to 1980's.\cite{s69,hnn85}
The basic ideas of the standard scenario are:

\noindent
1) disk hypothesis: A planetary system is formed from a
 protoplanetary disk with mass much less than its host star's mass, 

\noindent
2) planetesimal hypothesis: Building blocks of solid planets
 are km-sized rocky/icy ``planetesimals'' that are
 formed by coagulation of dust grains in the disk, 

\noindent
3) core accretion model: Gas components of gas giants are added
 after rocky/icy planets (cores) accrete from the planetesimals.

Though the standard scenario still has serious difficulties in formation
 of planetesimals from dust grains\cite{cy10}, discussion here starts
 with planetesimals, assuming that they have been formed by some
 mechanism.  
Planetesimals are bound by their self-gravity, but not by material
 strength.   
Except for the last stages of planet formation in which large planets
 perturb planetesimals, the relative velocities between planetesimals do
 not exceed the escape velocity from their surface, so collisions
 between them mostly result in accretion rather than bouncing or
 disruption.  
Since orbital damping timescale due to drag forces from disk gas and mean
 collision/close encounter timescale are much longer than orbital 
 periods, planetesimal motion is well described by the Kepler motion in 
 most of time.
The Keplerian orbital elements such as semimajor axis, orbital
 eccentricity, and orbital inclination are invariable in a two-body
 problem. 
These elements are changed impulsively by occasional collisions and
 close encounters and gradually by gas drag, although secular changes
 due to secular perturbations are sometimes important. 

Mass distribution of planetesimals evolves due to coagulation.
While the collisional cross-section is determined by the relative
 velocity, the relative velocity is regulated by the mass distribution.  
This complicated interplay leads to the non-linear planetesimal growth.  
Nevertheless, these processes are better understood than planetesimal
 formation from dust grains and gravitational interactions between a
 planet and a gas disk.

Since the sun is one of G dwarf stars that commonly exist in the Galaxy
 and planets are considered to be by-products of star formation, it was
 expected that planetary systems commonly exist around other stars in
 the Galaxy. 
Although searches for extrasolar planets (planets orbiting other stars) 
 started in 1940's, they ended without any detection or with false
 detection until the end of the last century. 
The first extrasolar planet around a main-sequence star, 51 Peg b, was
 eventually discovered in 1995, by radial velocity survey detecting
 wobble motion of the host star due to orbital motion of a
 planet.\cite{mq95}.
The discovered planet was a gas giant of $M \sim 10^2M_{\oplus}$,
 however, its orbital radius was as small as 0.05 AU and rotates around
 the host star in only 4 days. 
Such close-in giant planets are called ``hot jupiters.''
As of 2012, more than 750 extrasolar planets have been discovered.
In 2011, more than 2000 candidates have been identified by {\it Kepler}
 space telescope that detects eclipses of host stars by planets
 (transit survey).\cite{bea12}

Most of the extrasolar planetary systems discovered so far have 
 architectures quite different from the solar system.\cite{fea12}
Other than hot jupiters, many gas giants in eccentric orbits have been
 discovered. 
The origin of the diversity of planetary systems is now actively
 investigated.\cite{il08}
Both the radial velocity and transit surveys detect planets with
 relatively small orbital radii. 
Even with such restriction, the current surveys suggest that more than
 20\% of solar-type stars harbor close-in rocky/icy planets of 
 $M \lesssim 30M_\oplus$. 
The ubiquity of rocky/icy planets is another important subject.
It is inferred that rocky/icy planets commonly exist in habitable zones  
 that are the ranges of orbital radii where liquid water can exist on
 planetary surfaces.
Thus, possible detection of extrasolar life, in particular, biomarkers 
 on the extrasolar habitable planets through astronomical observations
 are also actively being discussed.\cite{ms10}

Here we mainly describe the basic dynamical and accretionary processes
 of planetesimals in the standard scenario of planet formation by showing
 $N$-body simulations.
Planetesimal accretion that is regulated by planetesimal dynamics is an
 important stage of planet formation since it determines the timescale
 of planet formation and the basic architecture of the planetary system.   
We also briefly summarize the dynamical effects of disk gas on planets
 and the core accretion model for formation of gas giants and discuss
 the diversity of planetary systems, extending and generalizing the
 model to extrasolar planetary systems.  
The model of protoplanetary disks is introduced in section
 \ref{section:disk_model}.  
We describe planetesimal dynamics and accretion in sections
 \ref{section:dyanmics} and \ref{section:accretion}.
The formation of terrestrial planets is presented in section
 \ref{section:terrestrial}. 
We introduce the current understanding of the planet-gas disk
 interaction and the core accretion model in section
 \ref{section:planet_disk}.   
Section \ref{section:summary} is devoted for a summary.
%

\section{Protoplanetary Disk Model}
\label{section:disk_model}

To investigate planet formation from a protoplanetary disk, we need a
 model of protoplanetary disks.
The standard disk model for solar system formation is the minimum-mass
 solar nebula (MMSN) model that is inferred from the mass distribution
 of planets in the present solar system in which the disk surface
 density (vertically integrated density) scales with the heliocentric
 distance as $r^{-3/2}$.\cite{h81}

Unfortunately current radio observations do not have enough resolutions
 to determine the surface density distribution of protoplanetary disks
 although a new large radio interferometer in Chile, ALMA, will
 bring us a lot of information on protoplanetary disks
 (e.g., Refs. \citen{wd05,cea11}).
Thus, we adopt a generalized disk model that is based on the MMSN model.
The surface density  distributions of dust (rock/ice) and gas components
 are given by 
\begin{eqnarray}
 \Sigma_\mathrm{d} & = &
  10 \epsilon_\mathrm{ice} f_\mathrm{d} 
  \left(\frac{r}{{\rm 1 AU}}\right)^{-3/2} {\rm gcm}^{-2},  
  \label{eq:sigma_dust} \\
 \Sigma_\mathrm{g} & = &
  2400 f_\mathrm{g} 
  \left(\frac{r}{{\rm 1 AU}}\right)^{-3/2} {\rm gcm}^{-2},  
  \label{eq:sigma_gas}
\end{eqnarray}
 where $f_\mathrm{d}$ and $f_\mathrm{g}$ are multiplicative factors to
 scale disk surface densities of dust and gas components.
The disk with $f_\mathrm{g} = f_\mathrm{d} = 1$  corresponds to a 50\%
 more massive MMSN model.

The step function $\epsilon_\mathrm{ice}$ is 1 inside the ice line $a_{\rm
ice}$ and 4.2 outside $a_{\rm ice}$.
The ice line is the location where the disk temperature coincides with
 the condensation temperature of H$_\mathrm{2}$O.
In this simple prescription, we set the ice line to that determined
 by an equilibrium temperature in optically thin disk regions\cite{h81} 
\begin{equation}
T = 
 280\left(\frac{r}{1 \mathrm{AU}}\right)^{-1/2}
 \left(\frac{L_*}{L_\odot}\right)^{1/4} \mathrm{K},
\end{equation}
 where $L_*$ and $L_{\odot}$ are the stellar and solar luminosity,
 as 
\begin{equation}
a_{\rm ice} = 2.7 \left(\frac{L_\ast}{L_\odot}\right)^{1/2} {\rm AU}.
\label{eq:a_ice}
\end{equation}
  
Note that the magnitude of $\epsilon_\mathrm{ice}$ may be modified by
 the local viscous dissipation and stellar irradiation. 
Although $\Sigma_\mathrm{d}$ and $\Sigma_\mathrm{g}$ are not necessarily
 proportional to $r^{-3/2}$ in general, we here present results with
 this dependence for simplicity (for different $r$-dependence models,
 see Ref.\citen{ki02}). 
The total mass of MMSN within radius $\simeq 30$ AU is 
 $\simeq 0.01M_{\odot}$ and radio observations show that the total disk
 mass around young stars is distributed in a range of
 0.001--$0.1M_{\odot}$.\cite{bs96}
So, the range of $f_\mathrm{d}$ and $f_\mathrm{g}$ may be 0.1--10.
Hereafter we call the disk model with $f_\mathrm{g} = f_\mathrm{d} = 1$ 
 as the standard disk model. 

The rotational angular velocity of gas taking into account the
 radial pressure gradient $\partial P/\partial r$ of disk gas is 
\begin{equation}
\Omega_\mathrm{g} = 
 \left(\frac{GM_\ast}{r^3} + \frac{1}{\rho r}
 \frac{\partial P}{\partial r} \right)^{1/2} 
 \simeq
 \Omega (1-\eta ), 
\end{equation}
 where
\begin{equation}
\eta \equiv 
 -\frac{1}{2} 
 \left(\frac{c_\mathrm{s}}{v_\mathrm{K}}\right)^2
 \frac{\partial \ln P}{\partial \ln r}
 \sim 
 \left(\frac{c_\mathrm{s}}{v_\mathrm{K}}\right)^2
 \sim 
 10^{-3} 
 \left(\frac{r}{1\mbox{AU}} \right)^{1/2}
 \left( \frac{M_*}{M_\odot} \right)^{-1}
 \left( \frac{L_*}{L_\odot} \right)^{1/4},
\label{eq:eta_pressure_grad}
\end{equation}
 $c_\mathrm{s}$ is the sound velocity of gas, $\Omega$ and
 $v_\mathrm{K}$ are the Kepler angular velocity and velocity, and we
 used $P = c_\mathrm{s}^2\rho$.\cite{h81}

%

\section{Planetesimal Dynamics} 
\label{section:dyanmics}


The timescale of planetesimal accretion is usually much longer than that
 of planetesimal dynamics in which planetesimal orbits evolve by mutual
 gravitational interaction.  
Thus planetesimal dynamics controls planetesimal accretion.
In this section we review the basic planetesimal dynamics due to two-body
 gravitational relaxation. 
The two basic effects of two-body gravitational relaxation in a
 planetesimal disk are viscous stirring and dynamical friction: viscous
 stirring increases the random velocity of planetesimals, while
 dynamical friction realizes the energy equiparation of the random
 energy.  
We also introduce the orbital repulsion of protoplanets.

\subsection{Two-body relaxation}

The equation of motion for a planetesimal is given as
\begin{equation}
 \frac{\mathrm{d}^2\mbox{\boldmath $x$}_i}{\mathrm{d}t^2} =
  -G M_* \frac{\mbox{\boldmath $x$}_i}{|\mbox{\boldmath $x$}_i|^3}
  + \sum_{j=1,j\neq i}^NGm_j\frac{\mbox{\boldmath $x$}_j-
  \mbox{\boldmath $x$}_i} 
  {|\mbox{\boldmath $x$}_j-\mbox{\boldmath $x$}_i|^3}
  + \mbox{\boldmath $F$}_\mathrm{gas}
  + \mbox{\boldmath $F$}_\mathrm{col},
\end{equation}
 where $m$ and $\mbox{\boldmath $x$}$ are the mass and position of
 planetesimals, $M_*$ is the stellar mass, and $G$ is the
 gravitational constant. 
The terms on the r.h.s. express from left to right the solar gravity,
 the mutual gravitational interaction of planetesimals, the force from
 disk gas, and the velocity change by collisions.  
We neglect the indirect term due to planetesimals since the total mass
 of planetesimals is much smaller than the solar mass.
For a planetesimal disk, the solar gravity is dominant except for the
 rare cases of close encounters of planetesimals and thus the orbit of
 planetesimals is almost Keplerian.  
The mutual gravitational interaction of planetesimals is the main
 perturbing force in a planetesimal disk.
The effect of gas will be discussed in section
 \ref{section:planet_disk}. 

The orbit of planetesimals is characterized by the semimajor axis $a$, 
 the eccentricity $e$, and the inclination $i$. 
The orbital eccentricity and inclination of planetesimals increase as a
 result of gravitational scattering among planetesimals. 
The deviation velocity, $v$, of a planetesimal from the Kepler
 velocity $v_\mathrm{K}$ of the local non-inclined circular orbit is
 called as random velocity, which is given by
\begin{equation}
 v \simeq (e^2+i^2)^{1/2}v_\mathrm{K}.
\end{equation}
The random velocity is an important factor that controls planet
 formation as will be shown later.
For example, the growth timescale of planetesimals depends on the
 random velocity as $t_\mathrm{grow}\propto v^2$ when gravitational
 focusing is effective in collisions.
Note that the thickness of a planetesimal disk is proportional to the 
 inclination (random velocity in a vertical direction) of planetesimals.   

The orbit of planetesimals gradually changes due to the mutual
 gravitational interaction.
This process is equivalent to the two-body relaxation process in star
 clusters.   
In terms of stellar dynamics, a planetesimal disk is a collisional
 system like globular clusters, in the sense that the system evolves
 through two-body encounters.  

The timescale of two-body relaxation for an equal-mass ($m$) many-body
 system is given by 
\begin{equation}
\label{eq:t_2b}
 t_\mathrm{relax} \equiv \frac{\sigma^2}{\mathrm{d}\sigma^2/\mathrm{d}t}
  \simeq \frac{1}{n \pi {r_\mathrm{g}}^2 \sigma\ln\Lambda} 
  = \frac{\sigma^3}{n \pi G^2m^2\ln\Lambda},
\end{equation}
 where $n$ is the number density of constituent particles, $\sigma$ is
 the velocity dispersion, $r_\mathrm{g}$ is the gravitational radius
 given by $r_\mathrm{g}=Gm/\sigma^2$, and $\ln\Lambda$ is the Coulomb
 logarithm.\cite{bt87}
The denominator $n \pi {r_\mathrm{g}}^2 \sigma$ means the number of
 close encounters with deflection angle $> 90^\circ$ per unit time and 
 $\ln\Lambda$ takes into account the effect of distant encounters.
On this timescale a particle forgets the initial orbit.

The two roles of two-body relaxation in a planetesimal disk are
 viscous stirring and dynamical friction.
In the following, we illustrate these processes by showing examples of
 $N$-body simulations.   
The initial conditions of planetesimals used below are summarized
 as follows: 1000 equal-mass ($m=10^{24}$ g) planetesimals are
 distributed in a ring of the radius $a=1$ AU with width
  $\Delta a=0.07$ AU. 
The surface density of the ring is consistent with the standard disk
 model. 
The initial distributions of eccentricities and inclinations are set by
 the Rayleigh distributions with dispersions 
 $\langle e^2\rangle^{1/2}=\langle i^2\rangle^{1/2}=2r_{\rm H}/a$, where
 $r_{\rm H}$ is the Hill (Roche) radius of planetesimals given by
\begin{equation}
 r_{\rm H} = \left(\frac{2m}{3M_*}\right)^{1/3}a.
\end{equation}
The Hill radius is the radius of the gravitational potential well of a
 body in a rotating frame.
Note that $\langle e^2\rangle^{1/2}$ and $\langle i^2\rangle^{1/2}$
 are proportional to the velocity dispersions in radial and vertical
 directions, respectively, and 
 $\sigma \simeq (\langle e^2\rangle + \langle i^2\rangle)^{1/2}v_\mathrm{K}$. 
We set $M_* = M_\odot$.

The basic gravitational relaxation processes such as gravitational
 scattering among planetesimals are scaled by the Hill units
 ($r_{\rm H}$ and $\Omega^{-1}$).\cite{i90}

\subsection{Viscous stirring}

\begin{figure}
 \centerline{
 \includegraphics[width=0.485\textwidth]{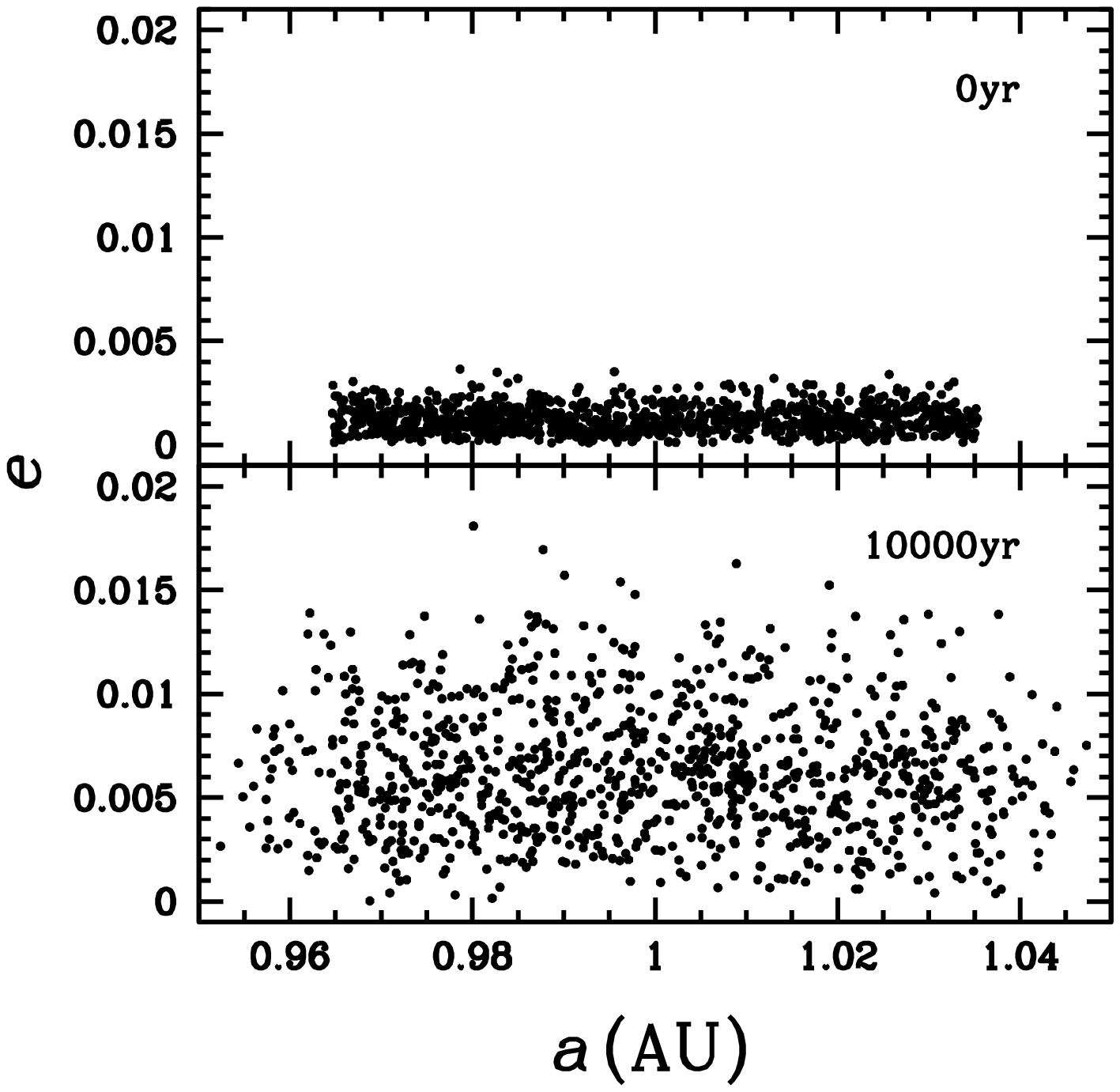}
 \hspace{1ex}
 \includegraphics[width=0.485\textwidth]{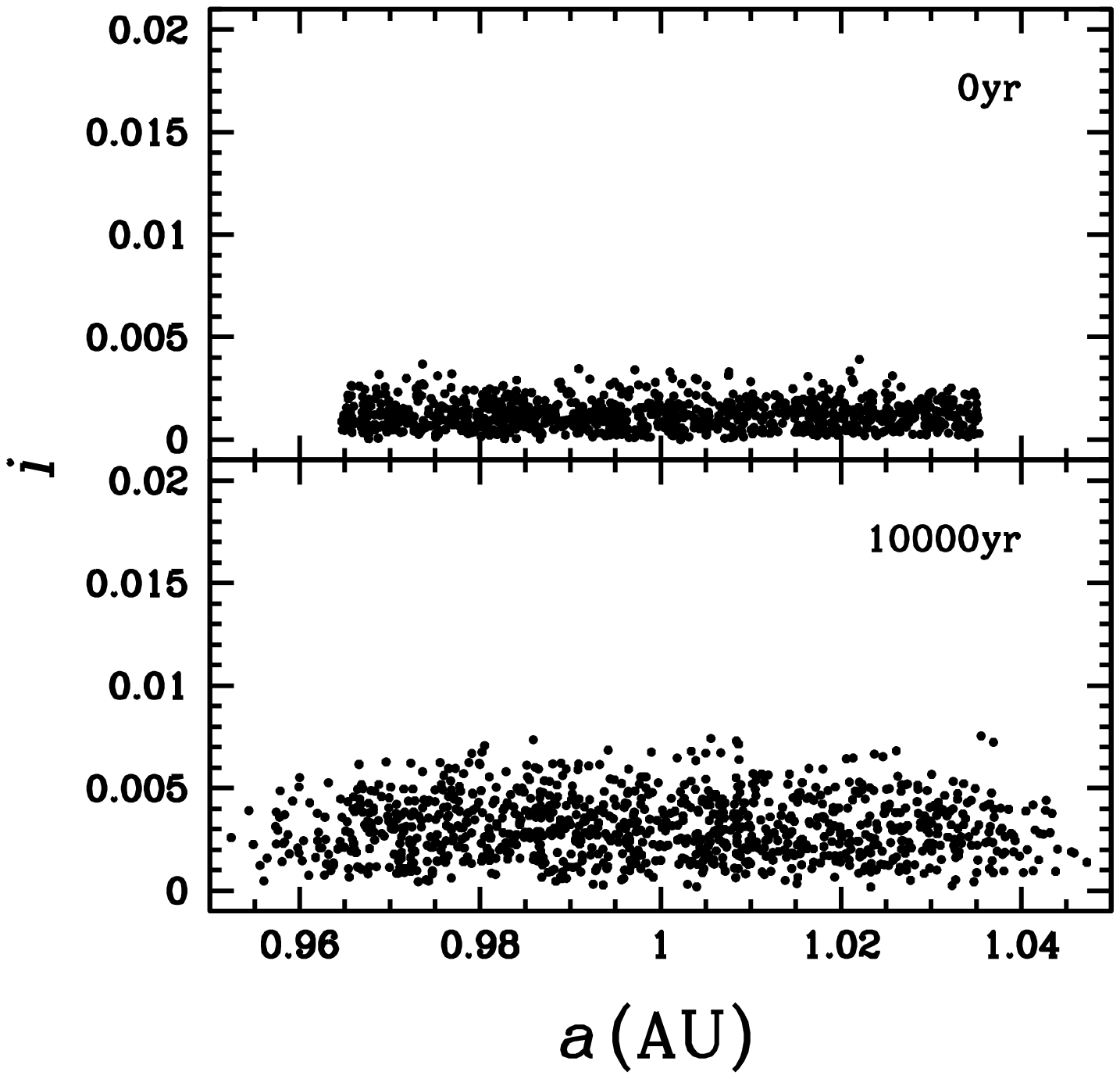}
 }
 \caption{Snapshots of the planetesimal system on the $a$-$e$ (left) and 
 $a$-$i$ (right) planes at $t=0$ (top) and 10000 year (bottom).}
 \label{fig:vs_ae}
\end{figure}

\begin{figure}
 \centerline{
 \includegraphics[width=0.485\textwidth]{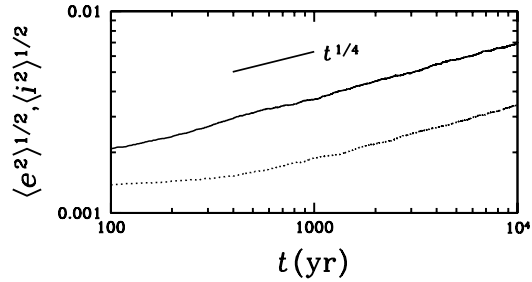}
 }
 \caption{Time evolution of $\langle e^2\rangle^{1/2}$ (solid) and
 $\langle i^2\rangle^{1/2}$ (dotted).} 
 \label{fig:vs_ts}
\end{figure}

The mutual gravitational interaction increases the random velocity of
 planetesimals on average. 
In other words, the Kepler shear velocity due to differential rotation
 is transfered to the random velocity. 
This process is called as viscous stirring.

Here we consider an equal-mass planetesimal disk for simplicity.
In the dispersion-dominated regime 
 ($\langle e^2\rangle^{1/2},\langle i^2\rangle^{1/2} \gtrsim r_{\rm H}/a
 $) where the relative velocity  
 of planetesimals is mainly determined by the random velocity, the
 viscous stirring rate for $\langle e^2\rangle$ is given by\cite{i90}
\begin{equation}
\label{eq:vs_e}
\frac{\mathrm{d}\langle e^2\rangle}{\mathrm{d}t} \sim 
\frac{\langle e^2\rangle}{t_{\rm relax}}.
\end{equation}
We can obtain the viscous stirring rate for $\langle i^2\rangle$ by
 replacing $e$ with $i$ in (\ref{eq:vs_e}).

Figures~\ref{fig:vs_ae} show the system snapshots at $t=0$ and 10000
 year on the $a$-$e$ and $a$-$i$ planes.  
The eccentricities and inclinations of most planetesimals significantly 
 increase in 10000 years. 
On average the increase of $e$ is larger than that of $i$.
The distributions of $e$ and $i$ relax into the Rayleigh distributions. 
We also see the diffusion of planetesimals in $a$, which is the result
 of random walk in $a$ due to two-body scattering.
These increases of $e$ and $i$ with the diffusion in $a$ are the basics
 of viscous stirring. 

Time evolutions of $\langle e^2\rangle^{1/2}$ and 
 $\langle i^2\rangle^{1/2}$ are shown in Figure~\ref{fig:vs_ts}. 
It is clearly shown that $\langle e^2\rangle^{1/2}$ and
 $\langle i^2\rangle^{1/2}$ increase with time as $t^{1/4}$, and
 $\langle e^2\rangle^{1/2}/\langle i^2\rangle^{1/2} \simeq 2$ that
 corresponds to the anisotropic velocity dispersions in radial and
 vertical directions. 
These properties of $\langle e^2\rangle^{1/2}$ and 
 $\langle i^2\rangle^{1/2}$ are the characteristics of two-body
 relaxation in a disk.\cite{ikm93} 

As the number density of planetesimals is inversely proportional to the
 disk thickness that is proportional to the velocity dispersion,
 $n\propto \sigma^{-1}$, we have $t_\mathrm{relax} \propto \sigma^4$. 
Thus, we have $\sigma \propto t^{1/4}$ from (\ref{eq:t_2b}).
The origin of the anisotropy of the velocity dispersion is the shear
 velocity between planetesimals with different $a$ due to the
 differential rotation in the Kepler potential.
This type of anisotropy is also known for the Galactic stellar
 disk.\cite{ki92} 


In reality the gas drag and collisions damp the random velocity of
 planetesimals. 
In planetesimal accretion, there exist equilibrium values of the
 eccentricity and inclination, at which viscous stirring and damping by
 the gas drag and collisions balance.

\subsection{Dynamical friction}
\label{section:df}

\begin{figure}[t]
 \centerline{
\includegraphics[width=0.485\textwidth]{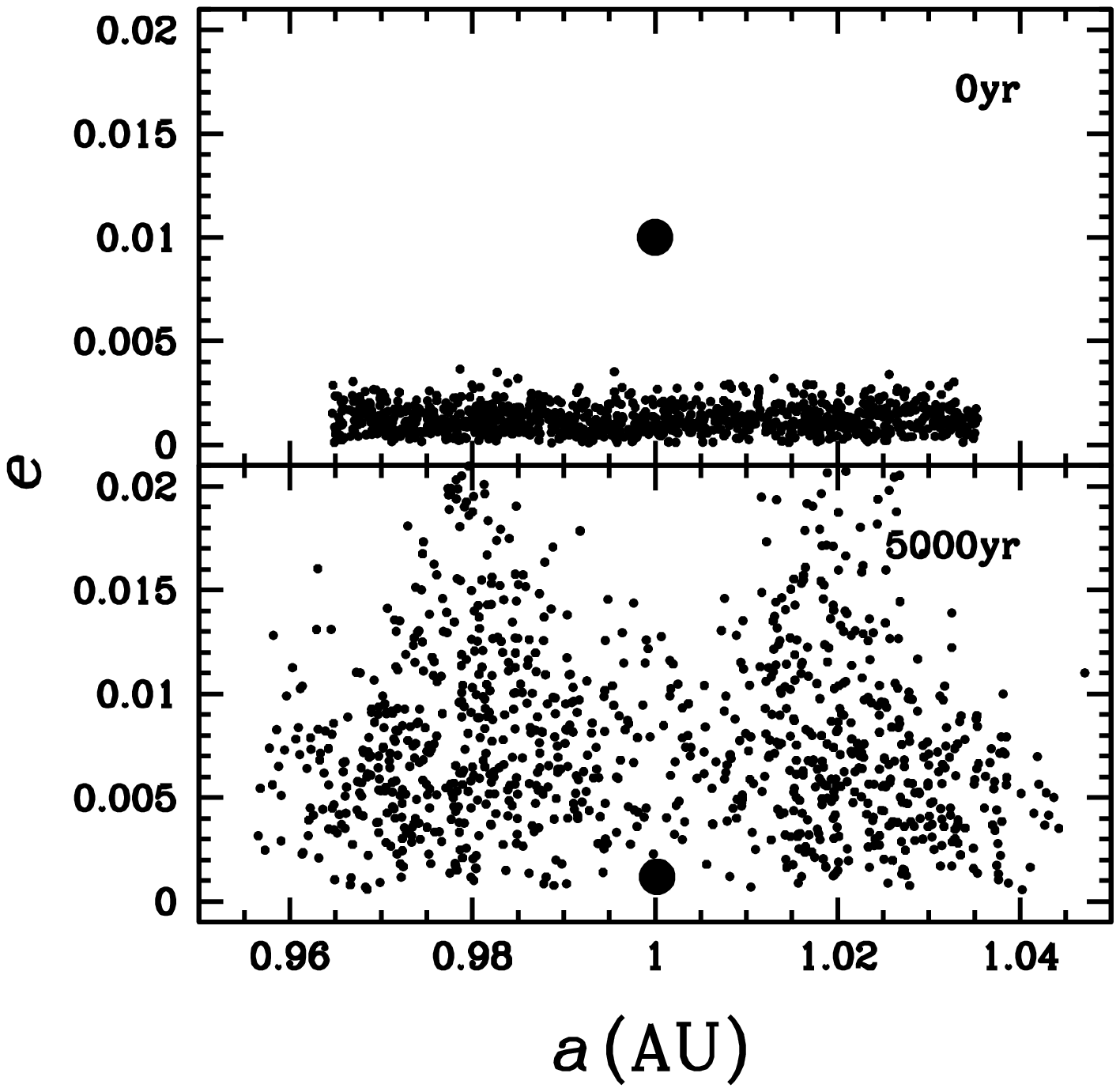}
 \hspace{1ex}
\includegraphics[width=0.485\textwidth]{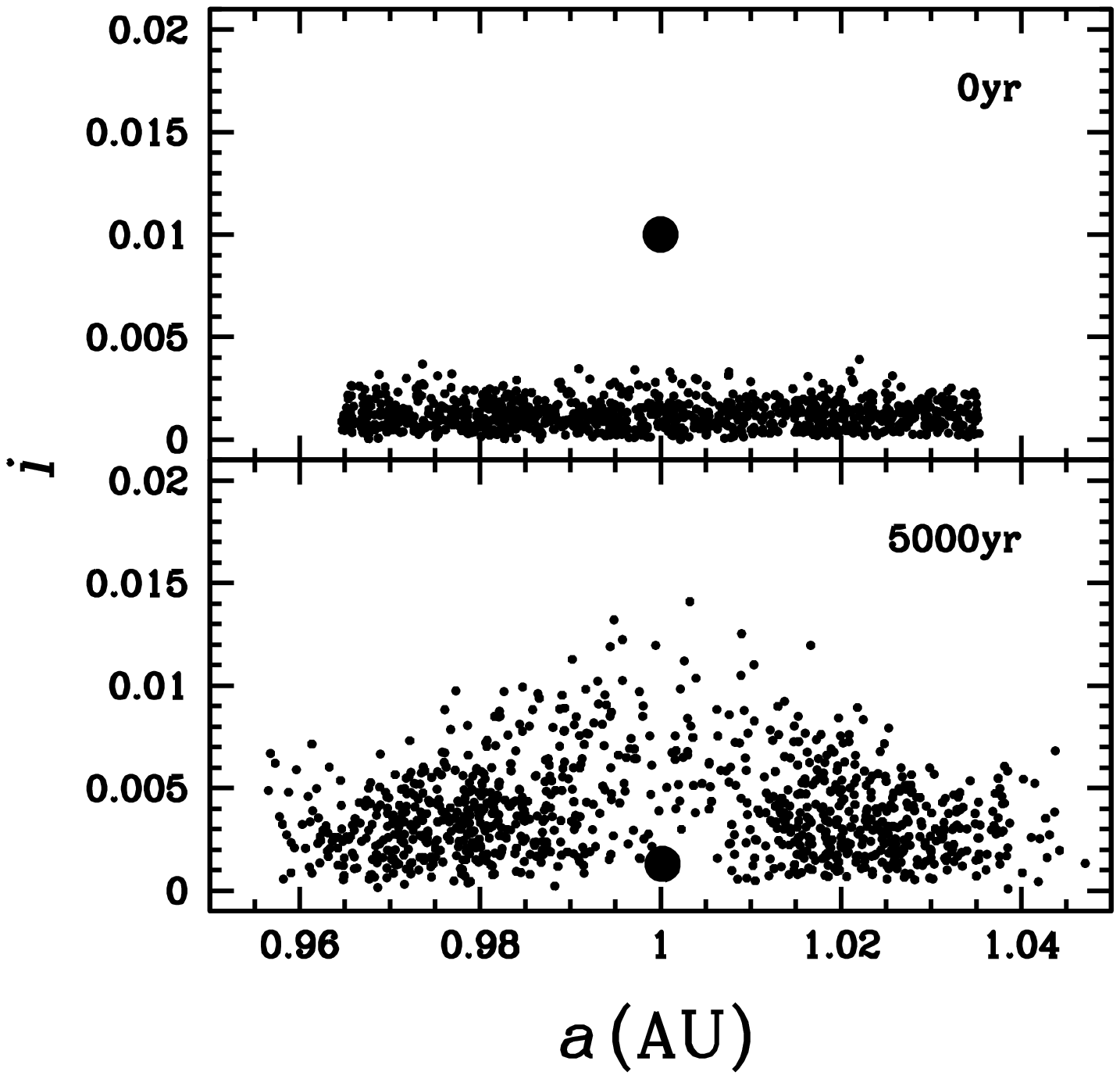}
 }
 \caption{Snapshots of the planetesimal system on the $a$-$e$ (left) and
 $a$-$i$ (right) planes at $t=0$ (top) and 3000 year (bottom). 
 The large circle indicates the protoplanet.}
 \label{fig:df_ae}
\end{figure}

\begin{figure}[t]
 \centerline{
 \includegraphics[width=0.485\textwidth]{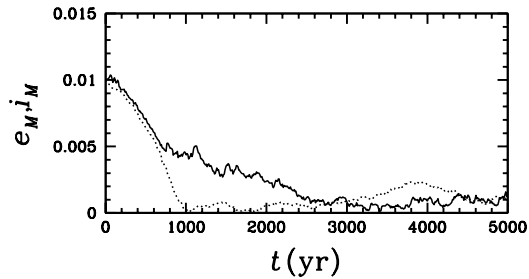}
 }
 \caption{
 Time evolution of $e_M$ (solid) and $i_M$ (dotted) of the
 protoplanet.
 } 
 \label{fig:df_ts}
\end{figure}

Next we consider the two-body relaxation for particles with a mass
 distribution. 
Dynamical friction is the process of the equiparation of the random
 energy, $(1/2)mv^2$, of planetesimals.
In other words, the random velocity becomes $v\propto m^{1/2}$.
As an illustration of dynamical friction, we consider a simple case with
 a protoplanet (large planetesimal) with $M=100m$ embedded in a swarm of
 planetesimals. 
We focus on the orbital evolution of the protoplanet.
The initial orbital elements of the protoplanet are $a_M=1$ AU and
 $e_M=i_M=0.01$. 

For $M\gg m$ and $e_M>e$, the dynamical friction rate for $e_M$ is
 approximated as\cite{i90}
\begin{equation}
\label{eq:df_e}
 \frac{\mathrm{d}e_M^2}{\mathrm{d}t} \sim 
 - \frac{m}{M}\frac{e_M^2}{t_{\rm relax}}. 
\end{equation}
The rate for $i_M$ is obtained by replacing $e_M$ with $i_M$ in
 (\ref{eq:df_e}). 
Figures~\ref{fig:df_ae} show the system snapshots at $t=0$ and 3000
 year on the $a$-$e$ and $a$-$i$ planes.  
We see that the eccentricity and inclination of the protoplanet decrease
 to almost 0 in 3000 years. 
However, its semimajor axis is kept almost constant.
On the other hand, the eccentricities and inclinations of the neighbor
 planetesimals of the protoplanet are raised by reaction.   
The V-like structure around the protoplanet on the $a$-$e$ plane
 corresponds to the constant Jacobi energy curve.
This heating of neighbor planetesimals by a protoplanet leads to the
 decrease of the growth rate of the protoplanet as will be shown in
 section \ref{section:oligarchic_growth}.\cite{im93}

Time evolution of $e_M$ and $i_M$ of the protoplanet is shown in
 Figure~\ref{fig:df_ts}. 
In 3000 years, the eccentricity and inclination of the protoplanet are
 reduced to $\sim 0.001$ due to dynamical friction from small
 planetesimals, in other words, the orbit of the protoplanet becomes a
 non-inclined nearly circular orbit.

One of the important features of dynamical friction is that the 
 dynamical friction rate does not depend on the mass of individual
 particles but the system density.
For protoplanet-planetesimal scattering, $t_{\rm relax}$ for
 the protoplanet is $t_{\rm relax} \propto (nM^2)^{-1}$ from
  (\ref{eq:t_2b}). 
Therefore, (\ref{eq:df_e}) leads to 
 $|\mathrm{d}e_M^2/\mathrm{d}t| \propto nm = \rho$, 
 where $\rho$ is the system density.

When the mass of a protoplanet is smaller than the total mass of
 planetesimals in its feeding zone, the recoil of planetesimal
 scattering damps $e_M$ and $i_M$ of the protoplanet as explained
 above. 
The recoil can also change the semimajor axis $a_M$ of the protoplanet
 under some circumstances.
We comment on it in section \ref{section:planet_disk}, after explaining
 orbital migration of a protoplanet due to interactions with a gas
 disk.

\subsection{Orbital repulsion} 
\label{section:repulsion}

\begin{figure}
 \centerline{
 \includegraphics[width=0.485\textwidth]{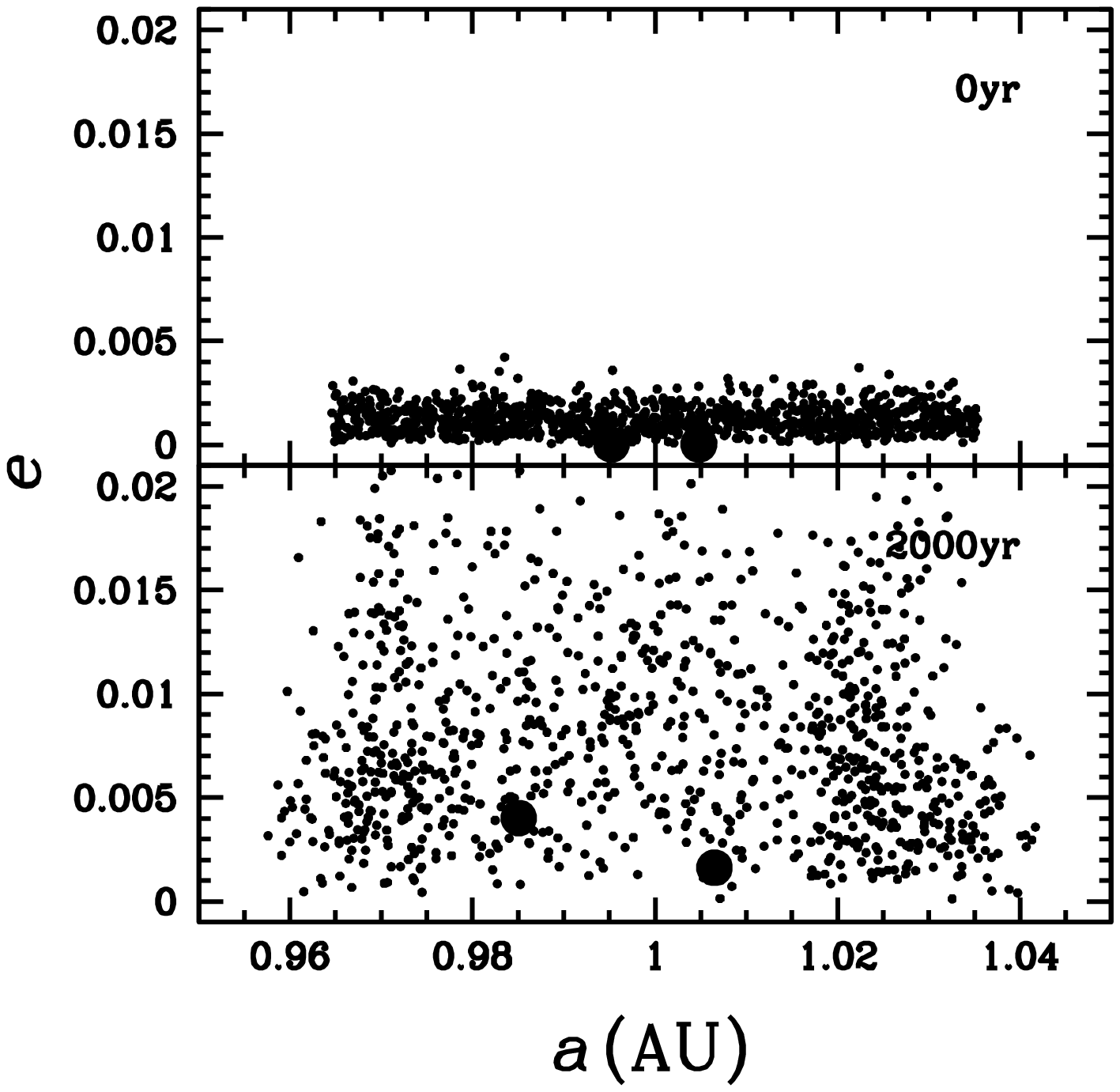}
 \hspace{1ex}
 \includegraphics[width=0.485\textwidth]{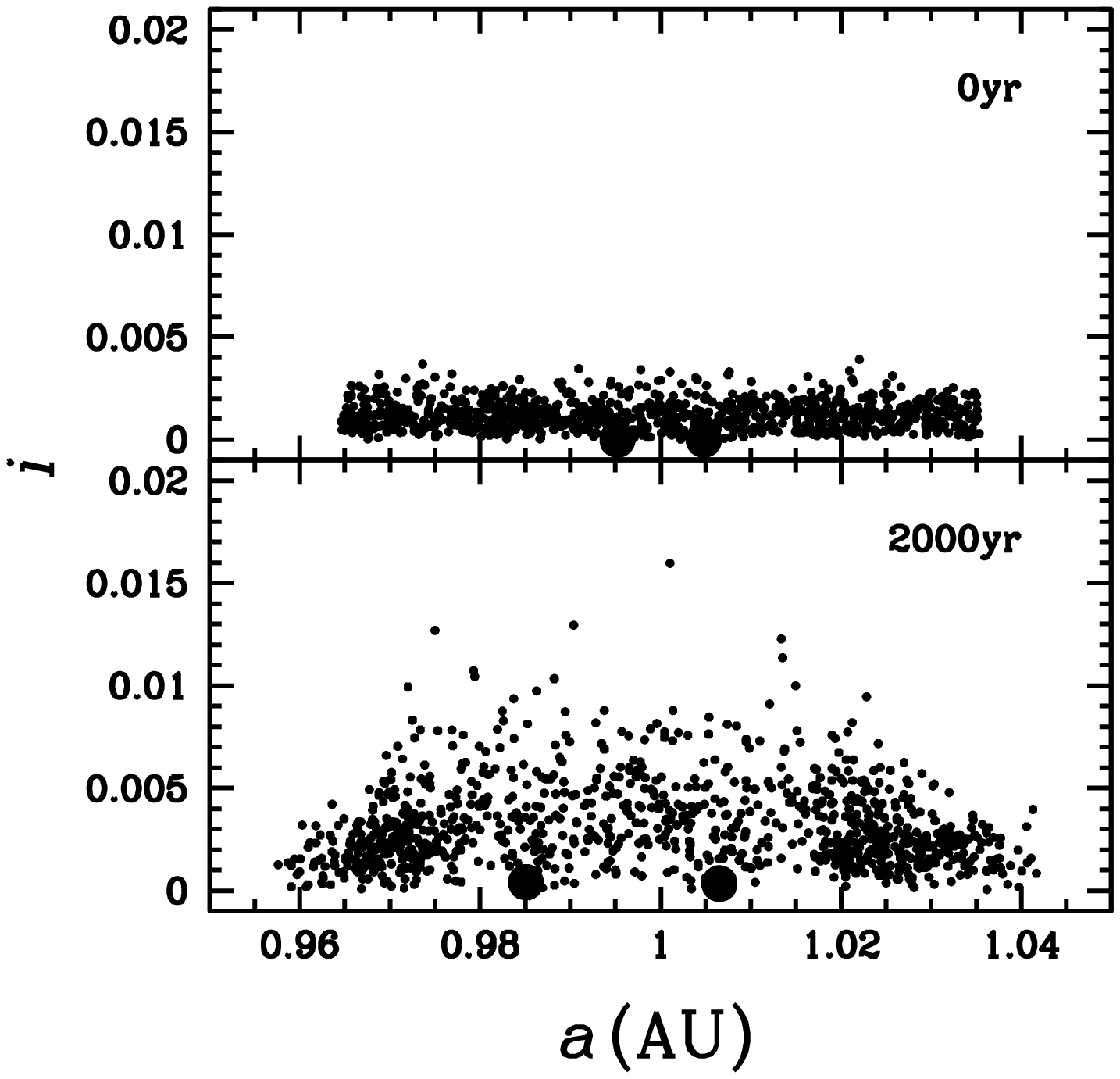}
 }
 \caption{Snapshots of the planetesimal system on the $a$-$e$ (left) and
 $a$-$i$ (right) planes at $t=0$ (top) and 2000 year (bottom).
 The large circles indicate the protoplanets.}
 \label{fig:or_ae}
\end{figure}

\begin{figure}
 \centerline{
 \includegraphics[width=0.485\textwidth]{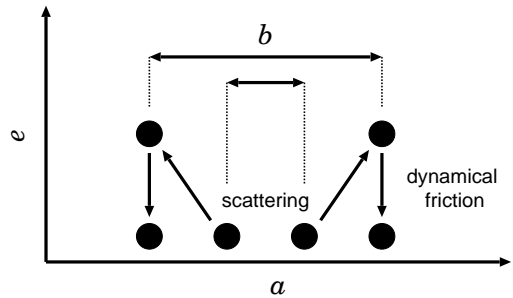}
 }
 \caption{Schematic illustration of orbital repulsion on the $a$-$e$ plane.} 
 \label{fig:repul}
\end{figure}

As an application of dynamical friction let us think the orbital
 evolution of two interacting protoplanets embedded in a swarm of
 planetesimals.  
It is found that if the orbital separation, $b$, of the
 two protoplanets is smaller than a few times their mutual Hill radius,
 they expand their orbital separation to $b\gtrsim 5r_{\rm H}$.\cite{ki95} 
This phenomenon is called as orbital repulsion.
In Figures~\ref{fig:or_ae}, the protoplanets with $M=100m$ initially on
 non-inclined circular orbits with the orbital separation of 
 $b=3r_{\rm H}$ expands the orbital separation to $b\simeq 8r_{\rm H}$ 
 keeping nearly non-inclined circular orbits in 2000 years.

The orbital repulsion is a coupling effect of gravitational scattering
 between large bodies and dynamical friction from small bodies.
Figure \ref{fig:repul} schematically explains the mechanism of orbital
 repulsion.
The mechanism consists of two stages: 
1) Scattering between two protoplanets on nearly circular non-inclined
 orbits increases their eccentricities and orbital separation. 
2) Dynamical friction from planetesimals reduces the
 eccentricities of the protoplanets while the orbital separation remains
 almost the same. 
Thus, the orbital separation of two protoplanets increases keeping
 nearly circular orbits. 
If the orbital separation is less than $5r_{\rm H}$, relatively strong
 scattering occurs and the separation expands rapidly.
We can analytically show the behavior of the first stage, based on the
 conservation of energy and angular momentum in two-body scattering
 under the solar gravity.

During the course of planetesimal accretion, as protoplanets grow, their
 orbital separation normalized by the Hill radius decreases, since
 $r_{\rm H}\propto M^{1/3}$.  
This implies that they repeat the orbital repulsion while growing. 
Consequently, the orbital separation of protoplanets is kept always
 larger than about $5r_{\rm H}$.  
This is one of the important factors that realize oligarchic growth of
 protoplanets. 
The typical orbital separation is $b \simeq 10r_{\rm H}$, which only
 weakly depends on the mass and the semimajor axis of protoplanets and  
 the disk surface density.\cite{ki98}

%

\section{Planetesimal Accretion} 
\label{section:accretion}


Orbiting the sun planetesimals sometimes collide with each other to form
 protoplanets.
In this section we review the basic physics of accretionary evolution of
 planetesimals.
Runaway growth of planetesimals and oligarchic growth of protoplanets
 are demonstrated by showing $N$-body simulations.

\subsection{Growth modes of planetesimals}
\label{section:growth_mode}


In general, there are two major modes when particles grow by
 coagulation, namely, ``orderly'' and ``runaway'' growth.
In the orderly growth mode, all the particles grow equally, in other
 words, mass ratios between particles tend to be unity.
On the other hand, in the runaway growth mode, larger particles grow
 faster than smaller ones and their mass ratios increase
 monotonically.
Which growth mode is relevant to planetesimal accretion had been
 controversial around the end of the last century.

Now we consider the evolution of mass ratio between two test particles 
 with mass $M_1$ and $M_2$ ($M_1>M_2$). 
The time derivative of the mass ratio is given by
\begin{equation}
\frac{\mathrm{d}}{\mathrm{d}t}\left(\frac{M_1}{M_2}\right) =
\frac{M_1}{M_2}
\left(
\frac{1}{M_1}\frac{\mathrm{d}M_1}{\mathrm{d}t} -
\frac{1}{M_2}\frac{\mathrm{d}M_2}{\mathrm{d}t} 
\right),
\end{equation}
 which shows that it is the relative growth rate
 $(1/M)(\mathrm{d}M/\mathrm{d}t)$ 
 that determines the growth mode of particles.
If the relative growth rate decreases with $M$,
 $\mathrm{d}(M_1/M_2)/\mathrm{d}t$ is negative and the mass ratio tends
 to be unity.
In this case, the growth mode is orderly.
On the other hand, if the relative growth rate increases with $M$,
 $\mathrm{d}(M_1/M_2)/\mathrm{d}t$ is positive and the mass ratio
 increases, which results in runaway growth. 

The growth rate of a test planetesimal with mass $M$ and radius $R$ by
 accreting field planetesimals with mass $m$ ($M>m$) is given by
\begin{equation}
\frac{\mathrm{d}M}{\mathrm{d}t} \simeq
n_m\pi R^2
\left(1+\frac{v_{\rm esc}^2}{v_{\rm rel}^2}\right)
v_{\rm rel}m,
\label{eq:growth_rate}
\end{equation}
 where $n_m$ is the number density of the field planetesimals, and
 $v_{\rm rel}$ and $v_{\rm esc}$ are the relative velocity between the
 test and the field planetesimals and the surface escape velocity from
 the test planetesimal, respectively.\cite{ki96}
For simplicity, we assume that all collisions lead to accretion
 (sticking probability is unity).  
Here $\pi R^2(1 + v_{\rm esc}^2/v_{\rm rel}^2)$ is the collisional
 cross-section with gravitational focusing, which can be easily obtained
 using the conservation of energy and angular momentum.
The term $v_{\rm esc}^2/v_{\rm rel}^2$ indicates the enhancement of
 geometrical collisional cross-section by gravitational focusing.
We discuss the growth mode of planetesimals using
 (\ref{eq:growth_rate}) below.

\subsection{Runaway growth of planetesimals}

\begin{figure}
 \centerline{
 \includegraphics[width=0.485\textwidth]{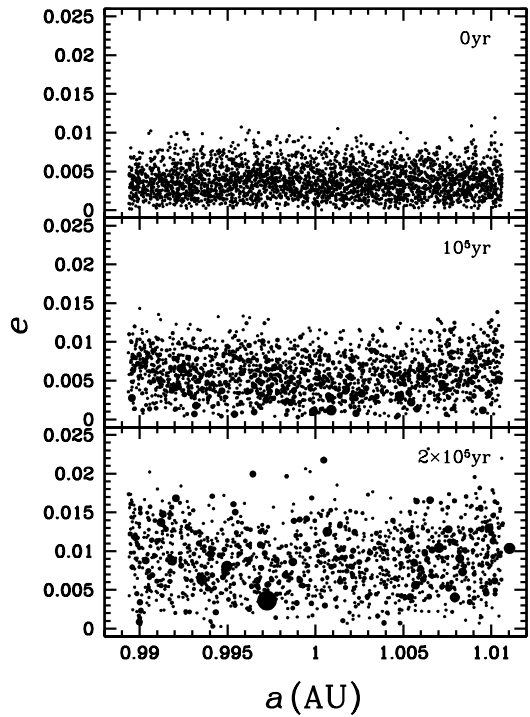}
 }
 \caption{
 Snapshots of the planetesimal system on the $a$-$e$ plane for $t=0$,
 $10^5$, and $2\times 10^5$ year.
 The circles represent planetesimals and their radii are proportional to
 the radii of planetesimals. 
 }
 \label{fig:a-e-t_runaway}
\end{figure}

\begin{figure}
 \centerline{
 \includegraphics[width=0.485\textwidth]{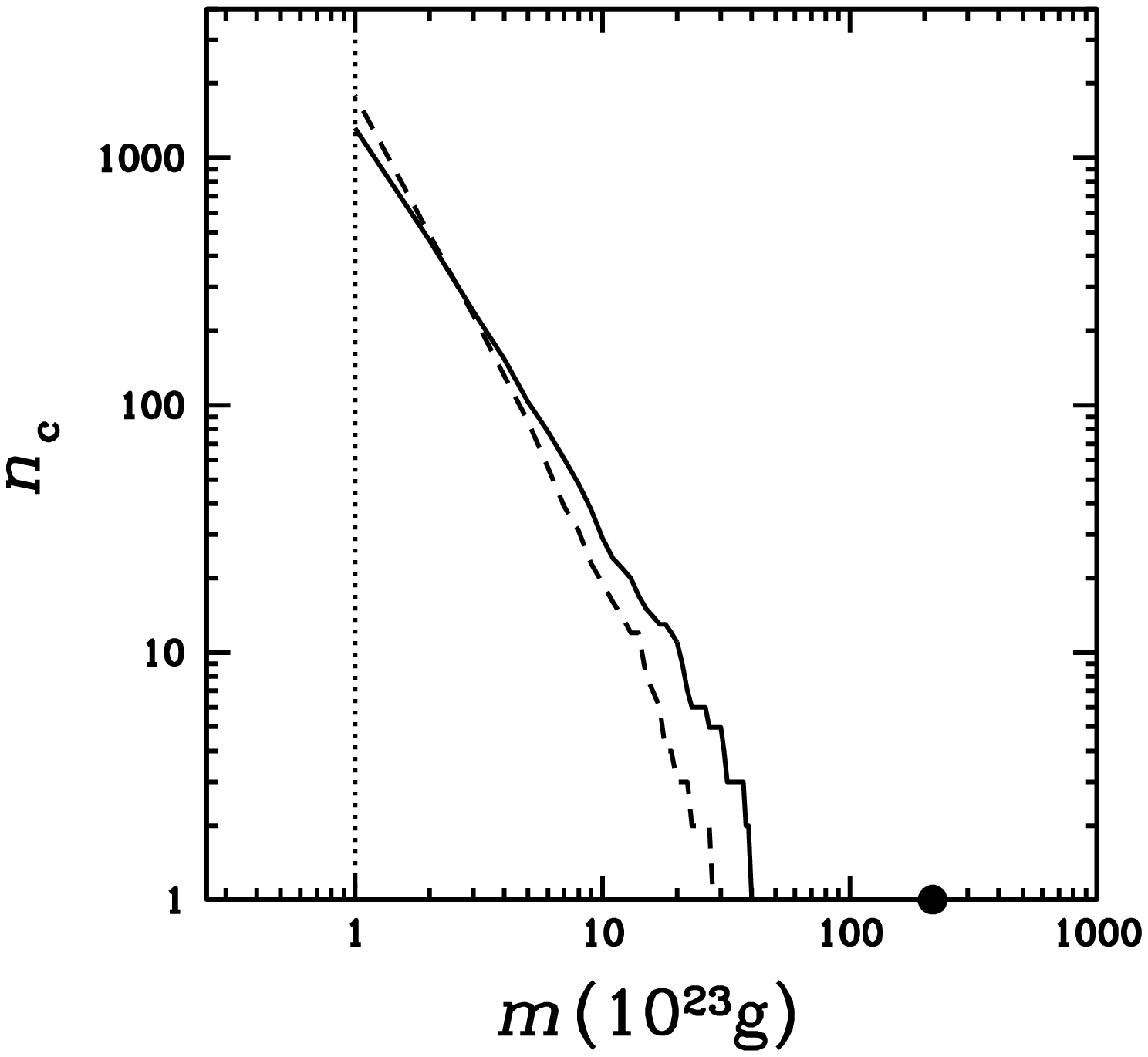}
 }
 \caption{
 Snapshots of the cumulative mass distribution $n_{\rm c}$ of the
 planetesimal system for $t=0$ (dotted), $10^5$ (dashed), and 
 $2\times 10^5$ (solid) year.
 The dot stands for a runaway body at $t=2\times 10^5$ year. 
 }
 \label{fig:m-n_c_runaway}
\end{figure}

In the early stages of planetesimal accretion, the growth mode of
 planetesimals is runaway growth, where larger planetesimals grow more
 rapidly than smaller ones and their mass ratios increase with time.
 \cite{gea78,ws89,ki96}
We illustrate the basic processes of runaway growth of planetesimals
 by showing the results of $N$-body simulation of planetesimal accretion.  
Figures \ref{fig:a-e-t_runaway} and \ref{fig:m-n_c_runaway} show an
 example of runaway growth.\cite{ki00}
Planetesimals are initially distributed in a ring around $a=1$ AU with 
 the width $\Delta a = 0.02$ AU.
The planetesimal system initially consists of 3000 equal-mass
 ($m=10^{23}$ g) bodies, which is consistent with the standard disk model.
The internal density of planetesimals is $\rho_\mathrm{p} = 2$
 gcm$^{-3}$. 
For simplicity, the calculation in this section is gas-free.

Figure \ref{fig:a-e-t_runaway} shows the system snapshots on the $a$-$e$
 plane for $t=0$, $10^5$, and $2\times 10^5$ year. 
In $2\times 10^5$ years, the number of bodies decreases to 1322.
It is clearly shown that a planetesimal grows in the runaway mode. 
At $t = 2\times 10^5$ year, the mass of the largest body reaches about
 200 times the initial mass, while the mean mass of planetesimals
 becomes only about twice larger. 
Note that the eccentricity and the inclination of the largest body is
 always kept small due to dynamical friction from smaller bodies.
These small eccentricity and inclination facilitate runaway
 growth.\cite{ws89,im92} 

The evolution of the mass distribution of planetesimals is shown in
 Figure~\ref{fig:m-n_c_runaway}.
The cumulative number of bodies $n_{\rm c}$ against mass is plotted.
The mass distribution relaxes to the distribution that is well
 approximated by a power-law distribution.
The runaway body at $t = 2\times 10^5$ year is shown by a dot that is
 separated from the continuous mass distribution. 
The mass range $10^{23}{\rm g} \lesssim m \lesssim 10^{24}{\rm g}$, which
 contains most of the system mass, can be approximated by
 $\mathrm{d}n_{\rm c}/\mathrm{d}m \propto m^{\alpha}$.
The power index calculated by using the least-square-fit method is
 $\alpha \simeq -2.5$.
This index can be derived analytically as a stationary
 distribution.\cite{mea98} 
The power index smaller than -2 is a characteristic of runaway growth,
 which means most of the system mass exists in small bodies.
Note that runaway growth does not necessarily mean that the growth
 time decreases with mass, but does mean that the mass ratio of any
 two bodies increases with time.
The runaway body keeps growing and then isolates from the continuous
 power-law mass distribution.
In this stage, the runaway body predominantly grows as a sink of the
 mass flow from the continuous power-law mass distribution. 

The runaway growth of planetesimals is explained as follows.
When gravitational focusing is effective,
 $v_{\rm esc}^2/v_{\rm rel}^2\gg 1$, (\ref{eq:growth_rate})
 reduces to  
\begin{equation}
\label{eq:rdmdt}
\frac{\mathrm{d}M}{\mathrm{d}t} \propto \Sigma_\mathrm{d} M^{4/3}v^{-2},
\end{equation}
 where $\Sigma_\mathrm{d}$ is the surface density of planetesimals and
 we used $n_m\propto \Sigma_\mathrm{d} v^{-1}$, 
 $v_{\rm esc} \propto M^{1/3}$, $R \propto M^{1/3}$, and 
 $v_{\rm rel}\simeq v$.
On the early stage of planetesimal accretion, $\Sigma_\mathrm{d}$ and
 $v$ barely depend on $M$, in other words, the reaction of growth on
 $\Sigma_\mathrm{d}$ and $v$ can be neglected since the mass in small
 planetesimals dominate the system.   
In this case we have 
\begin{equation}
\frac{1}{M}\frac{\mathrm{d}M}{\mathrm{d}t} \propto M^{1/3},
\end{equation}
 which leads to runaway growth as shown in section
 \ref{section:growth_mode}.

\subsection{Oligarchic growth of protoplanets}
\label{section:oligarchic_growth}

Protoplanets are formed through runaway growth of planetesimals.
In the late runaway stage, protoplanets grow while interacting with
 one another.
Kokubo and Ida\cite{ki98} investigated this stage and found oligarchic
 growth of protoplanets: similar-sized protoplanets grow keeping their 
 orbital separation larger than about $5r_{\rm H}$, while most
 planetesimals remain small.
Through oligarchic growth, a bi-modal protoplanet-planetesimal system
 is formed at the post-runaway stage.

\begin{figure}
 \centerline{
 \includegraphics[width=0.485\textwidth]{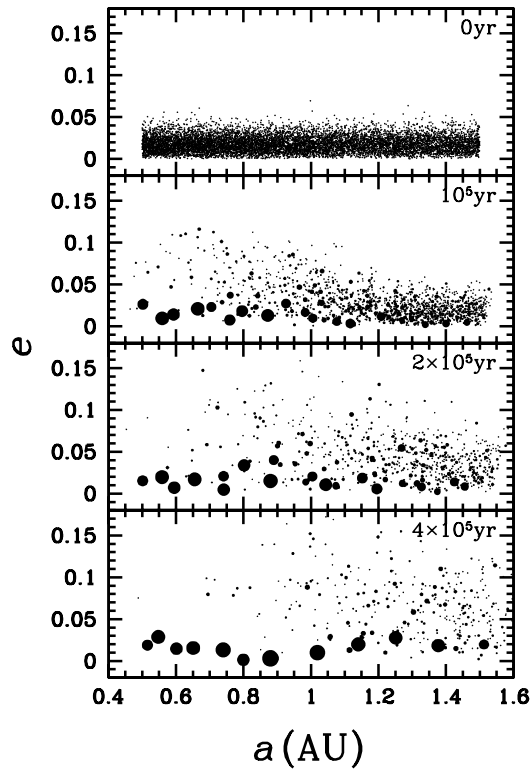}
 }
 \caption{
 Snapshots of the planetesimal system on the $a$-$e$ plane for $t=0$,
 $10^5$, $2\times 10^5$, and $4\times 10^5$ year.
 The circles represent planetesimals and their radii are proportional to
 the radii of planetesimals. 
 }
 \label{fig:a-e-t_oligarchic}
\end{figure}

We present an example of the $N$-body simulation that shows
 oligarchic growth.\cite{ki02}
Figure~\ref{fig:a-e-t_oligarchic} shows the result of a large scale
 simulation starting from $N=10000$ equal-mass ($m=1.5\times
 10^{24}$g) planetesimals in 
 $0.5\ \mathrm {AU} \leq  a \leq 1.5\ \mathrm{AU}$, which is consistent
 with the standard disk model. 
In this calculation, the 6-fold radius of planetesimals is used to
 accelerate the accretion process.
The use of the 6-fold radius of planetesimals does not change the
 growth mode of planetesimals but shorten the growth timescale about 6 
 times.\cite{ki96}

The accretion propagates from small to large $a$.
This is because the accretion timescale is smaller for smaller $a$ since 
 the surface number density of planetesimals is higher and the orbital 
 period is smaller for smaller $a$ [see (\ref{eq:t_grow})].
In $4\times 10^5$ years, the number of bodies decreases to 333.
About 10 protoplanets form with mass $\sim 10^{27}$ g on nearly
 circular non-inclined orbits with the orbital separation of
 $\simeq 10r_{\rm H}$. 
Note that at large $a$ the protoplanets is still growing.
The result of the $N$-body simulation is consistent with the estimation
 based on the oligarchic growth model described below. 

Oligarchic growth is the result of the self-limiting nature of runaway
 growth of protoplanets and orbital repulsion among protoplanets.
The formation of similar-sized protoplanets is explained by the
 slow-down of runaway growth.\cite{l87,im93}
Once the mass of the runaway body isolates from the continuous power-law
 mass distribution of planetesimals (see Figure~\ref{fig:m-n_c_runaway}),
 we can approximate the system by a two-component system: a protoplanet
 and small planetesimals. 
When the mass of the protoplanet $M$ exceeds about 100 times the typical 
 mass of planetesimals, the protoplanet effectively heats up the random
 velocity of neighbor planetesimals to be $v \propto M^{1/3}$.\cite{im93}
In this case, from (\ref{eq:rdmdt}), the relative growth rate
 becomes 
\begin{equation}
\frac{1}{M}\frac{\mathrm{d}M}{\mathrm{d}t} \propto \Sigma_\mathrm{d} M^{-1/3}.
\end{equation}
On this stage, $\Sigma_\mathrm{d}$ decreases through accretion of
 planetesimals by the protoplanet as $M$ increases.\cite{l87}
That is, the relative growth rate is a decrease function of $M$.
Thus the growth mode among protoplanets becomes orderly, in other words, 
 the mass ratios between protoplanets tend toward unity rather than
 increase.  
Thus, the neighboring protoplanets grow keeping similar masses.

Note that still in this stage, the mass ratio of a protoplanet to its
 neighbor planetesimals increases since for its neighbor planetesimals
 with mass $m$, 
 $(1/m)\mathrm{d}m/\mathrm{d}t \propto \Sigma_\mathrm{d} m^{1/3}M^{-2/3}$, 
 which leads to
\begin{equation}
\frac{(1/M)(\mathrm{d}M/\mathrm{d}t)}{(1/m)(\mathrm{d}m/\mathrm{d}t)} \propto \left(\frac{M}{m}\right)^{1/3}.
\end{equation}
The relative growth rate of the protoplanet is by a factor of
 $(M/m)^{1/3}$ larger than that of the planetesimals.
Thus, a bi-modal protoplanet-planetesimal system is formed.

While the protoplanets grow, orbital repulsion\cite{ki95}
 keeps their orbital separations $\simeq 10 r_{\rm H}$ as shown in
 section \ref{section:repulsion}.

\subsection{Isolation mass of protoplanets}
\label{section:isolation_mass}

As a result of the oligarchic growth of protoplanets, a protoplanet
 system is formed from a planetesimal disk.
The dynamical characteristics of the protoplanet system are estimated
 based on the oligarchic growth model.
In oligarchic growth, protoplanets are formed with a certain orbital
 separation.
Given this orbital separation $b$, the isolation (final) mass of a
 protoplanet at $a$ is estimated by\cite{ki02}
\begin{equation}
M_{\rm iso} \simeq 2\pi a b \Sigma_\mathrm{d} = 
0.16
f_\mathrm{d}^{3/2}
\epsilon_\mathrm{ice}^{3/2}
\left(\frac{b}{10r_{\rm H}}\right)^{3/2}
\left(\frac{a}{1{\rm AU}}\right)^{3/4}
\left(\frac{M_*}{M_\odot}\right)^{-1/2}
M_\oplus,
\label{eq:m_iso}
\end{equation}
 and the growth timescale of the protoplanet is
\begin{eqnarray}
 t_{\rm grow} & \equiv & \frac{M}{\mathrm{d}M/\mathrm{d}t} \nonumber \\ 
& \simeq & 
 1.3\times 10^5
 f_{\rm d}^{-1} 
 f_{\rm g}^{-2/5}
 \epsilon_\mathrm{ice}^{-1} 
 \left(\frac{M}{M_\oplus}\right)^{1/3} 
 \left(\frac{\rho_\mathrm{p}}{2 {\rm gcm}^{-3}}\right)^{3/5}
 \left(\frac{b}{10r_\mathrm{H}}\right)^{-2/5} 
 \left(\frac{a}{1{\rm AU}}\right)^{27/10} \nonumber \\
& &
 \left(\frac{m}{10^{18}{\rm g}}\right)^{2/15} 
 \left(\frac{M_*}{M_\odot}\right)^{-1/6} 
 \mathrm{years}.
\label{eq:t_grow}
\end{eqnarray}
Then the growth timescale of the protoplanet to the isolation mass is
 given by  
\begin{eqnarray}
\label{eq:t_grow_iso}
t_{\rm grow}(M_\mathrm{iso}) & \simeq &
0.7\times 10^5
f_\mathrm{d}^{-1/2}
f_\mathrm{g}^{-2/5}
\epsilon_\mathrm{ice}^{-1/2}
\left(\frac{\rho_\mathrm{p}}{2 {\rm gcm}^{-3}}\right)^{3/5}
\left(\frac{b}{10r_{\rm H}}\right)^{1/10}
\left(\frac{a}{1{\rm AU}}\right)^{59/20} \nonumber \\
& &
\left(\frac{m}{10^{18}{\rm g}}\right)^{2/15} 
\left(\frac{M_*}{M_\odot}\right)^{-1/3} 
\mathrm{years}.
\end{eqnarray}

\begin{figure}
 \centerline{
 \includegraphics[width=0.485\textwidth]{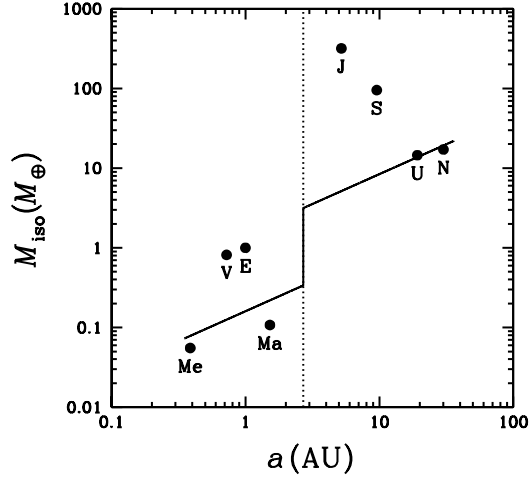}
 }
 \caption{
 Isolation mass of protoplanets against the semimajor axis with the mass
 of planets in the solar system and the ice line (dotted).
 }
 \label{fig:isolation_mass_ss}
\end{figure}

Figure~\ref{fig:isolation_mass_ss} shows the isolation mass of
 protoplanets against the semimajor axis for the standard disk model for
 solar system formation.\cite{ki00}
In the terrestrial planet region $M_\mathrm{iso} \sim 0.1 M_\oplus$,
 which is as large as the mass of Mercury and Mars.
This suggests that they are leftover protoplanets.
In order to complete Venus and Earth whose masses are one order of
 magnitude larger than that of protoplanets, further accretion of
 protoplanets is necessary. 
In the standard scenario of terrestrial planet formation the final stage
 is the ``giant impact'' stage of protoplanets.
The isolation mass of protoplanets in the gas giant region is around
 5--10$M_\oplus$, which may accrete gas from the disk before the
 dispersal of disk gas.
The mass ($\simeq 10M_\oplus$) and orbital separation ($\simeq 10$ AU)
 of ice giants are consistent with the oligarchic growth model.
However, their growth timescale is longer than the age of the solar
 system. 

The oligarchic growth model of protoplanets is now generally accepted as
 the standard process of planet formation though it still has some
 discrepancies.
The generalized oligarchic growth model is used to study the diversity
 of extrasolar planets together with the core accretion model as will be
 shown in section~\ref{section:formation_gas}.

%

%


\section{Terrestrial Planet Formation}
\label{section:terrestrial}


It is generally accepted that the final stage of terrestrial planet
 formation is the giant impact stage where protoplanets (planetary
 embryos) formed by oligarchic growth collide with one another to complete
 planets.\cite{w85,ki98}
This stage is being actively studied as many small extrasolar planets
 are discovered.

\subsection{Orbital instability of protoplanet systems}

\begin{figure}
 \centerline{
 \includegraphics[width=0.485\textwidth]{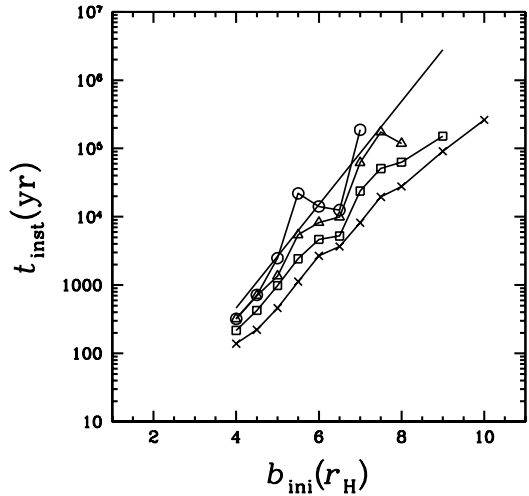}
 }
 \caption{
 Timescale of orbital instability against the initial orbital
 separation of protoplanets.
 The circles, triangles, squares and crosses indicate the initial 
$\langle e^2\rangle^{1/2} = 2\langle i^2\rangle^{1/2} = 0, 1h, 2h$, and
 4$h$, where $h$ is the reduced Hill radius $r_\mathrm{H}/a$. 
 The solid line shows the result of Chambers et al.\cite{cwb96}
 }
 \label{fig:b-t_inst}
\end{figure}

Through oligarchic growth protoplanets with orbital separation 
 $\simeq 10 r_\mathrm{H}$ are formed from planetesimals.
Though protoplanets perturb each other, the protoplanet system is
 orbitally stable when disk gas exists since its gravitational drag
 damps their eccentricities (see section~\ref{section:disk_interaction}).  
However observationally it is inferred that disk gas depletes on the
 timescale of 1-10 million years.\cite{bs96}
Thus in the long-term the protoplanet system becomes unstable through
 mutual gravitational perturbation after the dispersal of the gas disk. 

The timescale of the orbital instability of a protoplanet system is
 numerically obtained by $N$-body simulations as 
\begin{equation}
\log t_\mathrm{inst} \simeq
 c_1 \left(\frac{b_\mathrm{ini}}{r_\mathrm{H}}\right) + c_2, 
\end{equation}
 where $b_\mathrm{ini}$ is the initial orbital separation of
 adjacent protoplanets and $c_1$ and $c_2$ are functions of the initial
 $\langle e^2\rangle^{1/2}$ and $\langle i^2\rangle^{1/2}$ of the
 system.\cite{cwb96,ykm99} 
The physical interpretation of this semi-logarithm dependence is still
 unclear. 

Figure~\ref{fig:b-t_inst} shows the orbital instability timescale of a 
 protoplanet system consisting 10 equal-mass (0.1$M_\oplus$)
 protoplanets obtained by $N$-body simulations.\cite{ykm99}
Here ``instability'' means orbital crossing or collision of protoplanets.
The semi-logarithm dependence on the initial orbital separation of
 protoplanets is clearly shown. 
The constants $c_1$ and $c_2$ are smaller and larger for the larger
 initial $\langle e^2\rangle^{1/2}$ and $2\langle i^2\rangle^{1/2}$ of
 protoplanets, respectively. 
With only a small initial $\langle e^2\rangle^{1/2}$ and 
 $2\langle i^2\rangle^{1/2}$ the instability timescale 
 drastically shortens for large $b_\mathrm{ini}$.

\subsection{Giant impacts}

\begin{figure}
 \centerline{
 \includegraphics[width=0.485\textwidth]{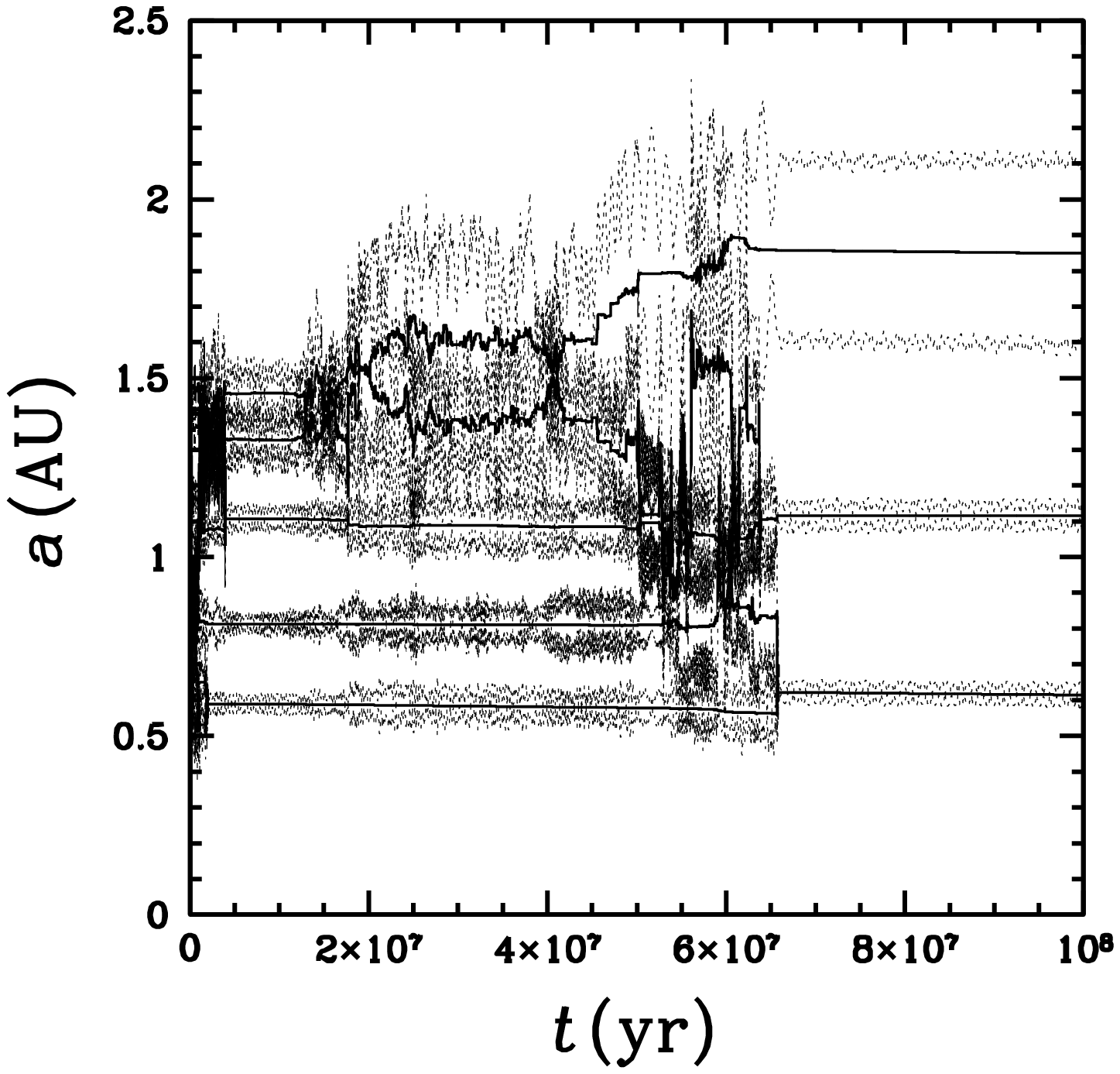}
 \hspace{1ex}
 \includegraphics[width=0.485\textwidth]{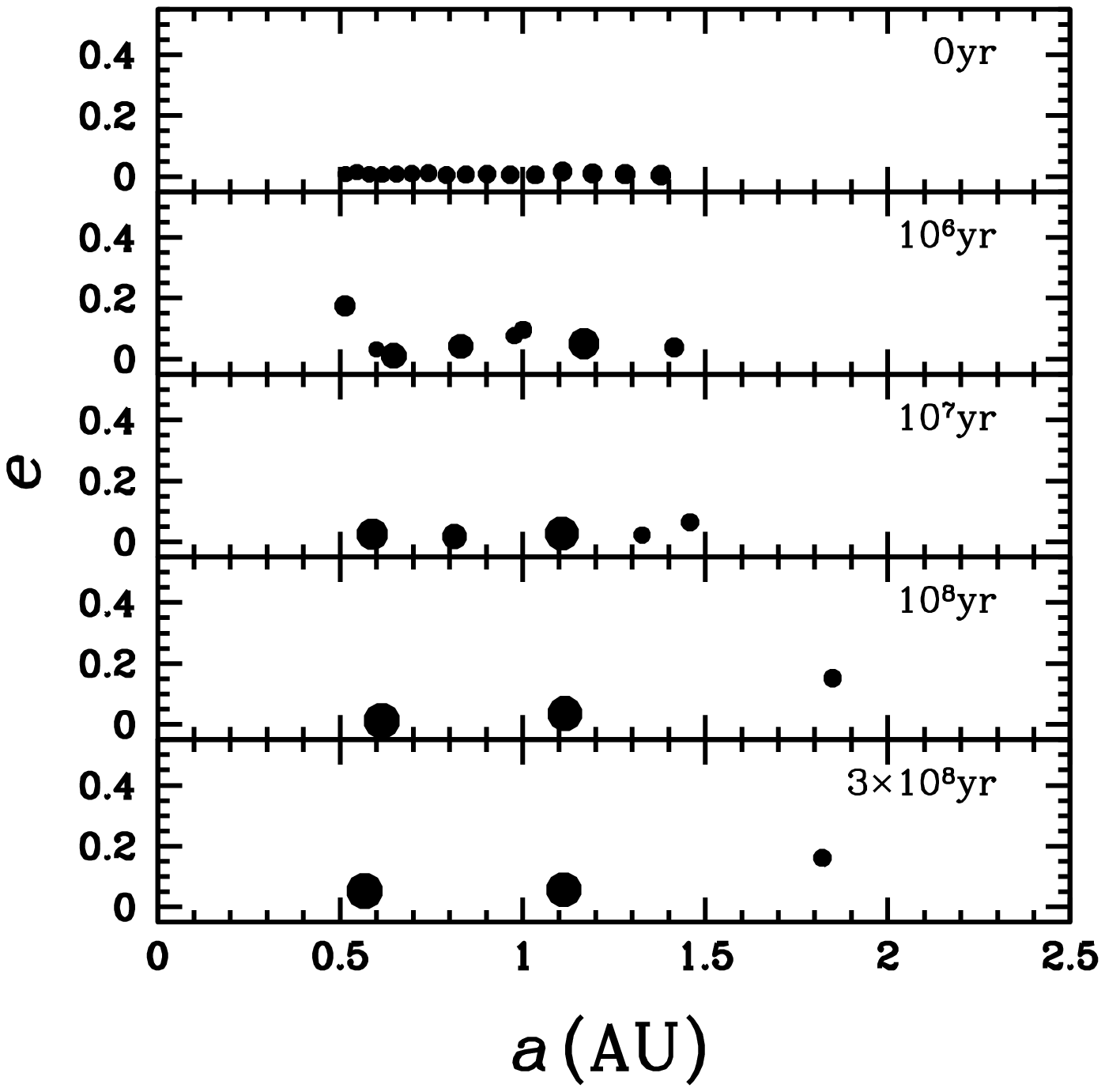}
 }
 \caption{
 Time evolution of the semimajor axes (solid lines) and pericenter and
 apocenter distances (dotted lines) of planets (left).
 Snapshots of the system on the $a$-$e$ plane at $t=0,10^6,10^7,10^8$,
 and $2\times 10^8$ year (right).
 The size of circles is proportional to the physical size of planets.
 }
 \label{fig:t-a_gi}
\end{figure}

After a protoplanet system becomes orbitally unstable the giant impact
 stage of protoplanets begins.
As this process is stochastic in nature, in order to clarify it, it is
 necessary to quantify it statistically.
Kokubo et al.\cite{kki06}, Kokubo and Ida\cite{ki07}, and Kokubo and
 Genda\cite{kg10} investigated the basic dynamics of the giant impact
 stage statistically with many $N$-body simulations.

\begin{figure}[t]
 \centerline{
 \includegraphics[width=0.485\textwidth]{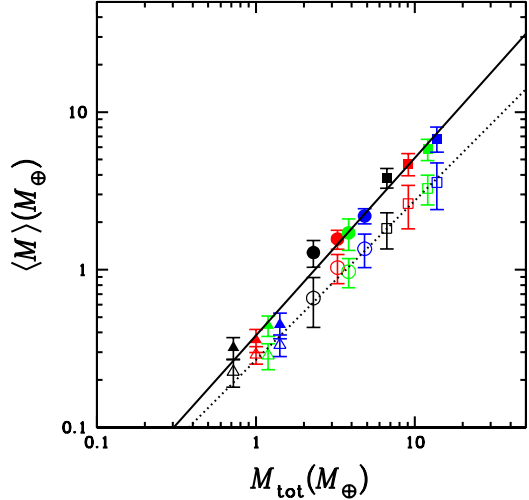}
 }
 \caption{
 Average masses of the largest $\langle M_1\rangle$ (filled symbols) and
 second largest $\langle M_2\rangle$ (open symbols) against $M_\mathrm{tot}$
 and the empirical fits for $\langle M_1\rangle$ (solid line) and
 $\langle M_2\rangle$ (dotted line).    
 The triangles, circles, and squares correspond to $f_\mathrm{d} = 0.3$, 1,
 and 3 models with $a_\mathrm{out} = 1.5, 2.0, 2.5$, and 3.0 AU.  
 }
 \label{fig:m_tot-m}
\end{figure}

Figure~\ref{fig:t-a_gi} shows an example where three terrestrial
 planets are formed from 16 protoplanets.\cite{kki06}
In this run the giant impact stage lasts for about $7\times 10^7$ years.   
In the standard disk model, two Earth-sized planets typically form in
 the terrestrial planet region. 
Kokubo et al.\cite{kki06} find that an important parameter of the initial
 protoplanet system for the number and mass of final planets is the
 total mass of protoplanets, $M_{\rm tot}$. 
The effects of the surface density distribution of the disk
 are unified using $M_{\rm tot}$.
In Figure~\ref{fig:m_tot-m}, the average masses of the largest planet
 $\langle M_1\rangle$ and the second-largest planet $\langle M_2\rangle$
 are plotted against $M_{\rm tot}$ for various models of protoplanet
 systems with different surface densities and radial extent together
 with their empirical fits.   
It is clearly shown that both $\langle M_1\rangle$ and 
 $\langle M_2\rangle$ increase almost linearly with $M_{\rm tot}$ and
 $\langle M_1\rangle \simeq 0.5M_{\rm tot}$ and 
 $\langle M_2\rangle \simeq 0.3M_{\rm tot}$.
This result shows that protoplanet accretion proceeds globally, in other
 words, over the whole terrestrial planet region.
Thus the large-scale radial mixing of material is expected.

\begin{figure}[t]
 \centerline{
 \includegraphics[width=0.485\textwidth]{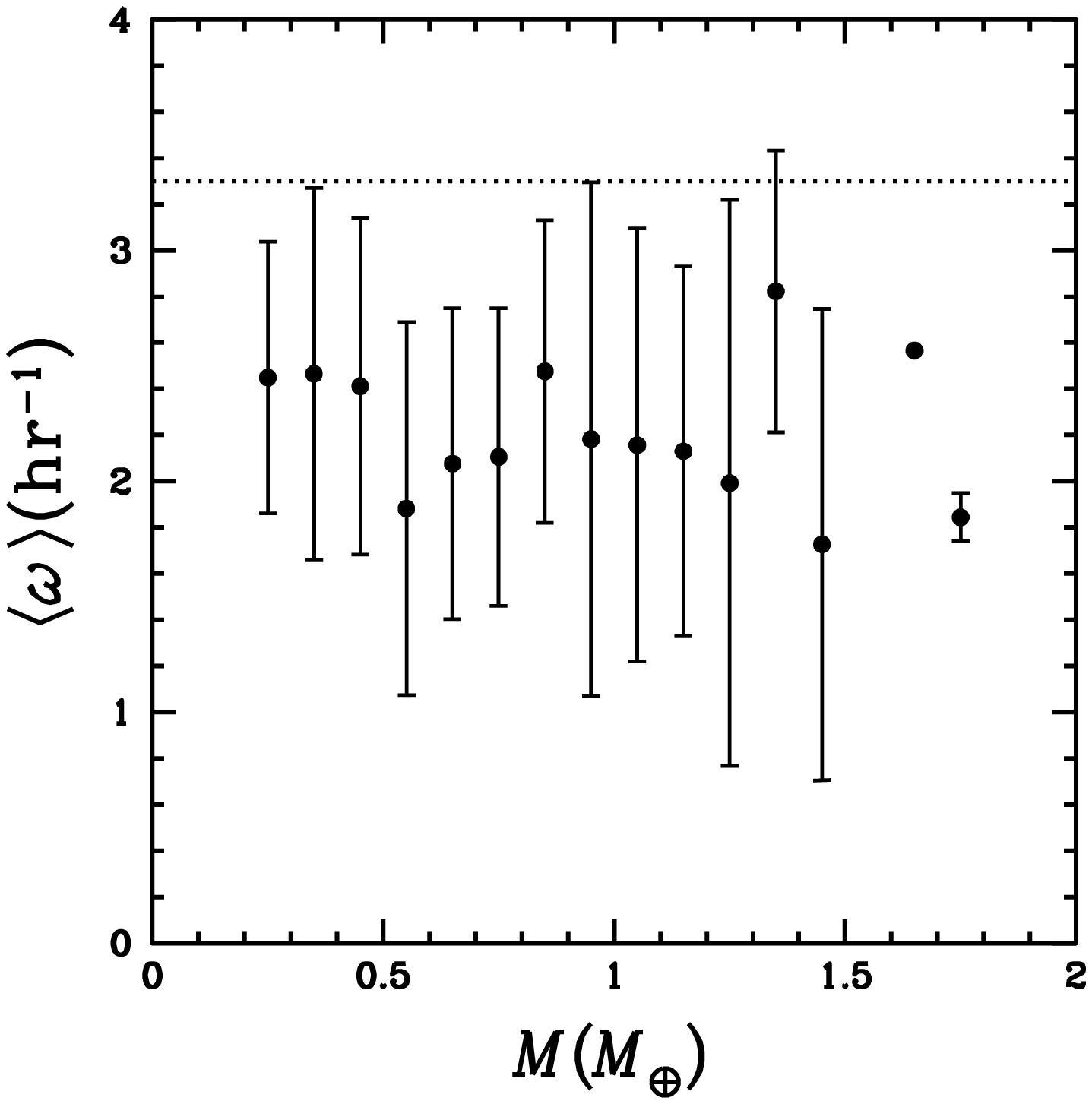}
 \hspace{1ex}
 \includegraphics[width=0.485\textwidth]{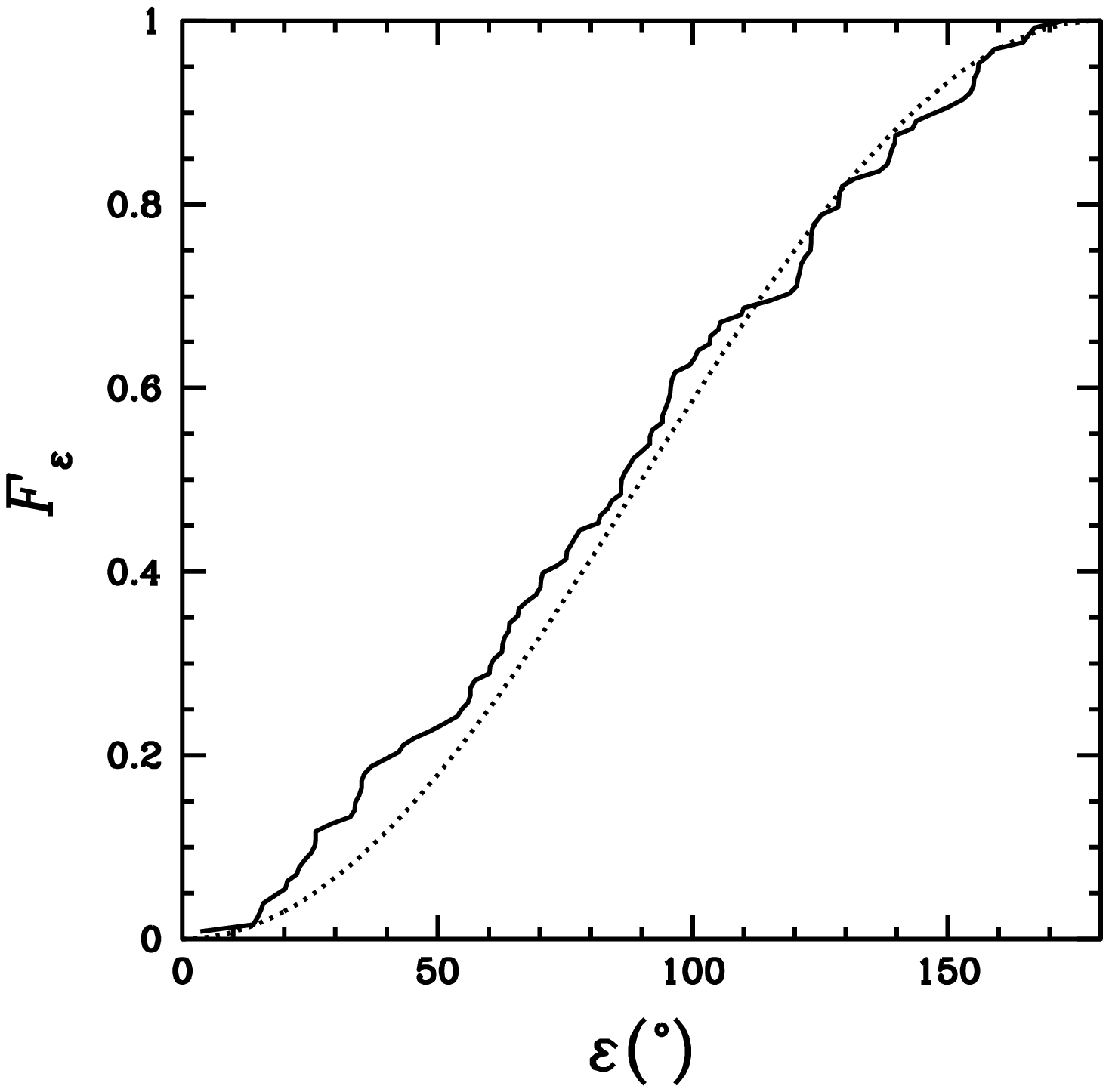}
 }
 \caption{
 Average spin angular velocity against planet mass (left) and
 normalized cumulative distribution of obliquity with the isotropic
 distribution (dotted line) (right) of all planets formed in the 50 runs
 of the standard model.  
 The error bars indicate the standard deviation and the dotted line shows
 $\omega_{\rm cr}$ (left).   
 }
 \label{fig:m-w_ave_std_all}
\end{figure}

The spin parameters of terrestrial planets are determined by the angular
 momentum brought by giant impacts.\cite{ki07}
The spin angular velocity averaged in mass bins against planet mass is
 shown in Figure~\ref{fig:m-w_ave_std_all}a.\cite{kg10}
Kokubo and Genda\cite{kg10} used the realistic accretion condition of
 protoplanets obtained by the numerical collision experiments of
 protoplanets.\cite{gki12}
It is clear that the average angular velocity is almost independent of mass. 
The average values are about 70\% of the critical angular velocity for
 rotational breakup 
\begin{equation}
\omega_{\rm cr} = \left(\frac{GM}{R^3}\right)^{1/2}.
\end{equation} 
This is a natural outcome for the giant impact stage where the impact
 velocity of protoplanets are $\sim v_\mathrm{esc}$.
For an Earth-mass planet, the spin angular velocity of
 $0.7\omega_\mathrm{cr}$ corresponds to the spin angular momentum of
 $9.5\times 10^{41}$ g cm$^2$ s$^{-1}$, which is 2.7 times larger than
 that of the Earth-Moon system.  
So the angular momentum of the Earth-Moon system is not a typical value
 but it is reasonably within the distribution of $\omega$.

In Figure~\ref{fig:m-w_ave_std_all}b, we show the obliquity distribution
 with the isotropic distribution
\begin{equation}
 n \mathrm{d}\varepsilon = 
 \frac{1}{2} \sin \varepsilon \mathrm{d}\varepsilon.    
\end{equation}
The obliquity ranges from 0$^\circ$ to 180$^\circ$ and
 follows an isotropic distribution.
The isotropic distribution of $\varepsilon$ is a natural outcome of
 giant impacts.
During the giant impact stage, the thickness of the protoplanet
 system is $\sim a\langle i^2\rangle^{1/2} \sim 10r_{\rm H}$, which is
 much larger than the size of protoplanets $R\sim 10^{-2}r_{\rm H}$,
 where $i$ is the inclination of protoplanets.
Thus, collisions are three-dimensional.
We find that these results are independent of the initial protoplanet
 system parameters. 

Our results clearly show that the distribution of spin obliquity is
 isotropic, which suggests that terrestrial planets formed through giant
 impacts are likely to have $\varepsilon \sim 90^\circ$, in other words,
 their spin axes are nearly on their orbital plane.
Prograde spin with small obliquity, which is common to terrestrial
 planets in the solar system except for Venus, is not a common feature
 for planets assembled by giant impacts. 
It should be noted, however, that the initial obliquity of a planet
 determined by giant impacts can be modified substantially by stellar
 tide if the planet is close to the star (Mercury) and by satellite tide
 if the planet has a large satellite (Earth).   

%

\section{Planet-Gas Disk Interaction} 
\label{section:planet_disk}


So far we have discussed dynamics and accretion of solid (rocky and
 icy) planets. 
In the solar system, gas giant planets, Jupiter and Saturn, exist.
Since the gas giants are 100 times or more massive than terrestrial
 planets, the architecture of planetary systems is sculpted by the
 giants, if they exist in the systems.  
The mass of protoplanetary disks is about one order of magnitude
 larger than that of giant planets, which implies that planetary
 orbits are significantly affected by interactions with disk gas.
Here we briefly summarize current understanding of planet-gas disk
 interactions and formation process of gas giant planets. 
Since distributions of gas giants in extrasolar planetary systems have
 been revealed by rapidly developing observations,  
 we can compare the theoretical predictions with the observed
 distributions. 
Although some aspects of the observed distributions are explained by the
 theory, there still remain important unsolved problems.

\subsection{Orbital evolution}
\label{section:disk_interaction}

While orbits of planetesimals and protoplanets can be eccentric,
 disk gas rotates in a circular orbit.
Thereby, planet-disk interaction generally damps the orbital
 eccentricity of planetesimals and protoplanets. 
For small planetesimals, aerodynamic gas drag is dominant because the
 ratio of the surface area to the volume is higher for smaller bodies.  
On the other hand, dynamical friction from disk gas is dominant for
 large bodies (protoplanets) that are more massive than the lunar mass
 ($\sim 0.01M_{\oplus}$). 

The damping timescale of the relative velocity $u$
 between a planetesimal (mass $M$ and physical radius $R$)
 and disk gas due to aerodynamical gas drag, in the case where 
 Reynolds number is larger than unity, is given by\cite{ahn76}
\begin{equation} 
t_{\rm stop} 
 \sim  \frac{M u}{\pi R^2 \rho_\mathrm{g} u^2}.
\label{eq:t_stop}
\end{equation}
The relative motion between a planetesimal/protoplanet and disk gas
 has two components: a component due to an eccentric orbit of the
 planetesimal ($\sim e v_{\rm K}$) 
 and an offset due to the pressure gradient in the gas 
 ($\sim \eta v_{\rm K}$).
Then, we have $u \sim (e+\eta)v_{\rm K}$.

On a timescale of $t_{\rm stop}$, 
both $e v_{\rm K}$ and $\eta v_{\rm K}$ are damped.
Thus, eccentricity damping timescale is
\begin{eqnarray}
t_{{\rm drag},e} & \equiv &
 \frac{e}{|\mathrm{d}e/\mathrm{d}t|} 
 \sim  t_{\rm stop} \nonumber \\
& \sim & 10^5 f_\mathrm{g}^{-1} \left(\frac{e+\eta}{0.1}\right)^{-1}
  \left(\frac{M}{M_{\oplus}}\right)^{1/3} 
              \left(\frac{\rho_\mathrm{p}}{3 \mbox{g cm}^{-3}}\right)^{2/3}
              \left(\frac{r}{1\mbox{AU}}\right)^{13/4}
              \mbox{years},
\label{eq:t_drag_e}
\end{eqnarray}
 where the gas density $\rho_{\rm g}$ is evaluated by $\Sigma_\mathrm{g}$ 
 divided by disk thickness $\sim 2 c_\mathrm{s}/\Omega$, and the sound
 velocity $c_\mathrm{s}$ is given by the equilibrium disk temperature
 $T$.  

The damping of $\eta v_{\rm K}$ causes inward radial migration of a
 planetesimal, because the planetesimal is subject to ``head wind'' to
 loose angular momentum, $L$.
Thus, $\mathrm{d}L/\mathrm{d}t \sim - M a \eta v_{\rm K}/t_{\rm stop}$.
Since $\mathrm{d}L/\mathrm{d}t \sim M \mathrm{d}(a v_{\rm
 K})/\mathrm{d}t = (1/2) M v_{\rm K} (\mathrm{d}a/\mathrm{d}t)$,  
 the migration timescale is given by
\begin{equation}
t_{{\rm drag},a} \equiv \frac{a}{|\mathrm{d}a/\mathrm{d}t|}  
\sim \frac{1}{2\eta} t_{\rm stop}.
\label{eq:t_drag_a}
\end{equation}
Equations~(\ref{eq:t_drag_e}) and (\ref{eq:t_drag_a}) show that 
 $t_{{\rm drag},a}$ is 2--3 orders of magnitude longer than 
 $t_{{\rm drag},e}$.  

The eccentricity damping timescale due to dynamical friction from disk
 gas is given by 
\begin{equation} 
t_{\rm df} \sim 
 \frac{M u}{\pi r_\mathrm{B}^2 \rho_\mathrm{g} u c_\mathrm{s}},
 \label{eq:t_df}
\end{equation}
 where $r_\mathrm{B} = GM/c_\mathrm{s}^2$ is the Bondi radius of the
 protoplanet.  
The last factor in the denominator is $c_\mathrm{s}$, which means that
 the velocity of density wave propagation is $\sim c_\mathrm{s}$ rather
 than $u$, since we consider the case of $c_\mathrm{s} > u$.
Note that the final form of $t_{\rm df}$ is similar to $t_{\rm relax}$
 (\ref{eq:t_2b}) if $\sigma$ is replaced by $c_\mathrm{s}$.
The eccentricity damping timescale is given by\cite{tw04}
\begin{eqnarray}
t_{{\rm df},e} & \equiv &
 \frac{e}{|\mathrm{d}e/\mathrm{d}t|} \sim
 t_{\rm df}
 \sim \frac{M_*}{M} \frac{M_*}{r^2 \Sigma_\mathrm{g}}
 \left(\frac{c_\mathrm{s}}{v_\mathrm{K}}\right)^{4}
  \Omega^{-1} \nonumber \\ 
& \sim & 
250 f_\mathrm{g}^{-1}
  \left(\frac{M}{M_{\oplus}} \right)^{-1}
  \left(\frac{r}{1{\rm AU}}\right)^{3/2}
  \left(\frac{M_*}{M_{\odot}}\right)^{3/2} {\rm years}.
  \label{eq:t_dfe}
\end{eqnarray}

Equations~(\ref{eq:t_stop}) and (\ref{eq:t_df}) show that
 the effects of gas drag and dynamical friction are 
 weaker for larger and smaller bodies, respectively, and
 $e$-damping is the weakest for $M \sim 0.01M_{\oplus}$ at 
 $r \sim 1$ AU.
The $e$-damping timescale even for bodies with $M \sim 0.01M_{\oplus}$ is
 shorter than the disk lifetime $\sim 1-10$ million years.
While the eccentricities of small planetesimals are excited by nearby
 protoplanets, allowing mutual collisions and accretion by the
 protoplanets, the eccentricities of the protoplanets are not excited,
 so that protoplanet-protoplanet collisions are generally inhibited in
 the presence of gas.  

Although planet-disk gravitational interactions are much more
 complicated than aerodynamical gas drag, the timescale of radial
 migration due to dynamical friction is evaluated in a similar way as  
\begin{equation}
t_{{\rm df},a} \sim \frac{1}{2\eta} t_{\rm df}.
\label{eq:t_migI}
\end{equation}
This migration is often called ``type-I migration.''
Tanaka et al.\cite{ttw02} derived a numerical factor 
 for (\ref{eq:t_migI}) through detailed 3-D linear calculation, 
 taking into a curvature effect (geometrical imbalance between the
 effects from inner and outer disks).
With the numerical factor, (\ref{eq:t_migI}) is
\begin{eqnarray}
t_{{\rm df},a} & = &
 0.23
 \frac{M_*}{M}
 \frac{M_*}{r^2 \Sigma_\mathrm{g}}
 \left(\frac{c_\mathrm{s}}{v_\mathrm{K}}\right)^{2}
 \Omega^{-1} \nonumber \\
 & \simeq &
  5 \times 10^4 
  f_\mathrm{g}^{-1}
  \left(\frac{M}{M_{\oplus}} \right)^{-1}
  \left(\frac{r}{1{\rm AU}}\right)^{3/2}
  \left(\frac{M_*}{M_{\odot}}\right)^{3/2} {\rm years}.
\label{eq:tau_mig1}
\end{eqnarray}
The migration timescales for an Earth-mass planet at 1AU
 and a core of $\sim 10 M_{\oplus}$ at 5 AU are $\sim 10^5$ years,
 which are 10--100 times shorter than disk lifetime, suggesting that
 the Earth and Jupiter's core cannot survive.
This is one of the most serious difficulties in planet formation theory
 today.
On the other hand, discovery of many close-in planets in extrasolar
 planetary systems requires this kind of inward migration. 
Many possibilities to retard, halt or reverse the migration are
 discussed, although the mechanism for planets to survive type I
 migration has not been identified yet.    

Orbital migration of protoplanets due to planetesimal scattering
 (gravitational interactions with a planetesimal disk) also exists, as
 pointed out in section \ref{section:df}.  
Although there is no systematic effect like $\eta$ in the gas disk,
 a curvature effect still exists.
While $\Sigma_\mathrm{d} \ll \Sigma_\mathrm{g}$, $c_\mathrm{s}$ is
 generally larger than the velocity dispersion of planetesimals, so that
 ``planetesimal-driven migration'' cannot be always neglected compared
 with type I migration.\cite{oit12}
However, the migration speed and direction have not been clarified yet.

\subsection{Formation of gas giants}
\label{section:formation_gas}

For the formation of gas giant planets, two models have been proposed.
In the ``core accretion'' model, a rocky/icy core first accretes
 from planetesimals and gas accretion onto the core follows.\cite{m80,bp86}
In the ``disk instability'' model, on the other hand, a protoplanetary
 disk fragments due to gravitational instability to directly form a gas
 giant(s).\cite{b97}
Since rocky/icy planets cannot be formed by the disk instability and
 even Jupiter and Saturn are metal-rich compared to the solar composition,
 the core accretion model has been regarded as a ``standard'' model.
Here, we summarize the core accretion model.
Note that a possibility of disk instability is now being revisited, because 
 extrasolar gas giants with large orbital radii, which are not
 easy to be formed {\it in situ} through the core accretion, 
 are being observed by direct imaging.

When the core mass becomes larger than a critical core mass, 
 pressure gradient no more supports envelope gas hydrodynamically
 against the core's gravity and hydrostatic envelope no longer exists.
The critical core mass ($M_{\rm c,cr}$) depends on planetesimal
 accretion rate onto the core ($\dot{M}_{\rm c}$) and the grain opacity
 in the envelope ($\kappa_{\rm gr}$). 
Through 1-D calculation, Ikoma et al.\cite{ine00} found that 
\begin{equation}
 M_{\rm c,cr} \simeq 
  10 \left( \frac{\dot{M}_{\rm c}}
      {10^{-6}M_{\oplus} {\rm  yr}^{-1}}\right)^{0.2\mbox{-}0.3} 
  \left( \frac{\kappa_{\rm gr}}{\kappa_{\rm gr,P}}\right)^{0.2\mbox{-}0.3}
  M_{\oplus},
\label{eq:crit_core_mass}
\end{equation}
 where $\kappa_{\rm gr,P}$ $(\sim 1{\rm cm^2 g^{-1}})$ is the grain
 opacity given by Pollack et al.\cite{pmc85}, who assumed dust grains with
 interstellar abundance and size distributions. 
Faster accretion and higher opacity (relatively large $\dot{M}_{\rm c}$
 and $\kappa_{\rm gr}$) result in a warmer planetary envelope and
 enhanced pressure gradient, so that $M_{\rm c,cr}$
 increases.\cite{s82,ine00} 

After $M_{\rm c}$ exceeds $M_{\rm c,cr}$, heat generation due to
 gravitational contraction of gas envelope itself supports the envelope
 against dynamical collapse and the envelope undergoes quasi-static
 contraction.\cite{bp86}
The contraction allows disk gas to flow from the protoplanetary disk
 into the Bondi radius of the planet, so that the contraction rate
 determines the rate of gas accretion onto the planet. 

Although the rate of the quasi-static contraction of envelope should be
 evaluated by radiative transfer calculation, we here present a simple
 estimate for the rate.\cite{si08} 
For $M_{\rm c} \sim M_{\rm c,cr}$, heat generation due to planetesimal
 accretion marginally equilibrates with the gravity of the core.
In the quasi-static envelope contraction, heat generation due to the
 envelope contraction marginally equilibrates with the gravity of the
 planet.  
Replacing $M_{\rm c}$ and $\dot{M}_{\rm c}$ in
 (\ref{eq:crit_core_mass}) by $M$ (total mass of the core and
 envelope) and $M/t_{\rm g,acc}$,  
 the gas accretion timescale $t_{\rm g,acc}$ is given by  
\begin{equation}
 t_{\rm g,acc} \simeq 
  10^7 \left( \frac{M}{10M_{\oplus}}\right)^{-({\rm 2.3\mbox{-}4})}
  \left( \frac{\kappa_{\rm gr}}{\kappa_{\rm gr,P}}\right) \; {\rm years}.
  \label{eq:tau_g,acc}
\end{equation}
Detailed radiative transfer simulations\cite{ine00,ig06} showed
consistent results. 

Since $t_{\rm g,acc}$ rapidly decreases with increasing $M$,
 the gas accretion is a runaway process and initial contraction
 regulates the total gas accretion timescale.
Equation (\ref{eq:tau_g,acc}) shows that for gas accretion to start the
 runaway gas accretion within disk lifetime ($t_{\rm dep}$) of 1-10
 million years, the core mass more than a few to ten Earth masses is
 required for $\kappa_{\rm gr} \sim (0.01\mbox{-}1) \kappa_{\rm gr,P}$.

\subsection{Formation sites of cores}

Using the results in section \ref{section:isolation_mass} and
 \ref{section:formation_gas}, we can quantitatively determine the
 formation sites of cores with a few $M_{\oplus}$ or more, neglecting
 type-I migration, which is highly uncertain. 
We consider a conservative condition for formation of gas giants
 such that $M_{\rm c} > 10 M_{\oplus}$.
In the following we use $b = 10 r_\mathrm{H}$, 
 $\rho_\mathrm{p} = 2.0$ g cm$^{-3}$, and $m = 10^{18}$ g.

In the inner regions, cores acquire the isolation mass $M_{\rm iso}$
 given by (\ref{eq:m_iso}), which shows that  
 $M_{\rm iso} > 10 M_{\oplus}$ only at $r > a_{\rm in}$ where 
\begin{equation}
 a_{\rm in} \simeq
 \left\{
 \begin{array}{ll}
 \displaystyle
  2.5\left(\frac{f_{\rm d}}{10}\right)^{-2}
 \left( \frac{M_*}{M_{\odot}} \right)^{2/3} \mbox{AU} & 
 \displaystyle
 \left[f_{\rm d} \gtrsim 10\left(\frac{M_*}{M_\odot}\right)^{-2/3}\right] \\
 \displaystyle
 a_{\rm ice} = 2.7 \left( \frac{M_*}{M_{\odot}} \right)^{2} \mbox{AU} & 
 \displaystyle
 \left [2\left(\frac{M_*}{M_\odot}\right)^{-2/3} \lesssim f_{\rm d} \lesssim
 10\left(\frac{M_*}{M_\odot}\right)^{-2/3}\right] \\
 \displaystyle
 3.5 \left(\frac{f_{\rm d}}{2}\right)^{-2} 
 \left( \frac{M_*}{M_{\odot}} \right)^{2/3} \mbox{AU} & 
 \displaystyle
 \left[f_{\rm d} \lesssim 2\left(\frac{M_*}{M_\odot}\right)^{-2/3}\right] 
 \end{array}
 \right.,
\label{eq:a_in}
\end{equation}
 where we assumed $L_* \propto M_*^4$.
This equation shows that preferred locations of cores for giant planets
 are the regions beyond the ice line ($r > a_{\rm ice}$), 
 although formation of large cores is possible even inside the ice
 line for the most massive disks with $f_{\rm d} \gtrsim 10$.   

On the other hand, in outer regions, core growth is so slow that the
 core mass does not reach the isolation mass before the depletion of
 disk gas.
Then, the condition for formation of cores with $M_{\rm c}$ larger than
 $10 M_{\oplus}$ is $t_{\rm grow}(10 M_{\oplus}) < t_{\rm dep}$, where
 we assume $t_{\rm dep} \sim 10$ million years and $t_\mathrm{grow}$ is
 given by (\ref{eq:t_grow}).
This condition imposes an outer limit of the semimajor axis
 ($a_{\rm out}$) for formation of gas giants:
\begin{equation}
a_{\rm out} \simeq
 6.4 f_{\rm d}^{14/27}
\left(\frac{\epsilon_\mathrm{ice}}{4.2}\right)^{10/27} 
\left(\frac{M_*}{M_\odot}\right)^{5/81} 
\left(\frac{t_{\rm dep}}{10^7 {\rm years}}\right)^{-10/27} {\rm AU},
\label{eq:a_out}
\end{equation}
 where solar abundance, $f_{\rm d} = f_{\rm g}$, is assumed.
 Formation of cores for gas giants is
 possible in the regions with $a_{\rm in} < r < a_{\rm out}$.

\subsection{Diversity of planetary systems}

\begin{figure}
 \centerline{
 \includegraphics[width=0.485\textwidth]{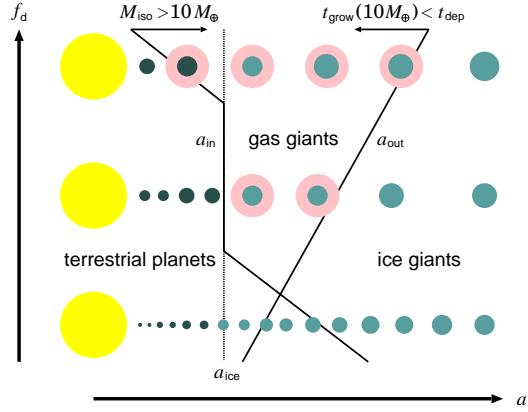}
 }
 \caption{
 Schematic illustration of diversity of planetary systems.
 }
 \label{fig:diversity}
\end{figure}

Using the results in the last subsection, we discuss the diversity of 
 planetary systems.\cite{ki02,il08} (Figure~\ref{fig:diversity}).
We now know that $M \gtrsim 10 M_\oplus$ within $t_\mathrm{dep}$ only at
 $a_{\rm in} < r < a_{\rm out}$.  
In low-mass disks for which $a_{\rm in} > a_{\rm out}$, gas giants are
 not formed. 
Equations (\ref{eq:a_in}) and (\ref{eq:a_out}) suggest that formation of
 gas giants is not expected in disks with $f_{\rm d} \lesssim 1$ (disks
 with masses smaller than the mass of the standard disk).
On the other hand, in high-mass disks, it is expected that multiple gas
 giants are formed since the range between $a_{\rm in}$ and
  $a_{\rm out}$ is broad. 
The multiple giants often undergo close scattering among them,
 resulting in formation of gas giants in eccentric orbits that are found
 in extrasolar planetary systems.
In massive disks, cores and gas giants form early enough when full
 amount of disk gas still exists. 
Then, the gas giant that has opened up a gap in the gas disk would
 migrate together with disk accretion (type II
 migration)\cite{lp85,lp93} to the vicinity of the host star. 
Thus, initial disk masses would produce diversity of planetary
 systems.\cite{il08}
 
Gas accretion rate onto the core and the mechanism to stop gas accretion
 are being discussed by many authors, as well as realistic type I
 migration rate. 
Although these processes sill have large uncertainties, the planetesimal
 dynamics and accretion described here are relatively well understood and
 they are fundamental pieces for discussion on formation of planetary
 systems.

%

\section{Summary}
\label{section:summary}

In the standard scenario of planet formation terrestrial and icy planets
 and cores of gaseous planets are formed by accretion of planetesimals. 
The random velocity (eccentricity $e$ and inclination $i$) of
 planetesimals controls planetesimal accretion and then growth of
 planetesimals changes the random velocity of planetesimals. 
This interplay of planetesimal dynamics and accretion show interesting
 phenomena of planetesimal accretion.
We have demonstrated the basic dynamical and accretionary processes of
 planetesimals by showing the examples of $N$-body simulations.  

In orbital evolution of planetesimals two-body gravitational relaxation
 plays key roles. 
The important basic processes are viscous stirring and dynamical friction. 
Viscous stirring increases dispersions $\langle e^2 \rangle^{1/2}$ and
 $\langle i^2 \rangle^{1/2}$ in proportion to time as  $t^{1/4}$, keeping
 the ratio $\langle e^2 \rangle^{1/2}/\langle i^2 \rangle^{1/2} \simeq 2$, 
 which is the characteristics of two-body relaxation in a Keplerian disk.  
Dynamical friction realizes the equiparation of the random energy, in
 other words, $e,i \propto m^{-1/2}$.
Orbital repulsion of two protoplanets embedded in a swarm of
 planetesimals keeps the orbital separation of the two protoplanets
 $b \gtrsim 5r_{\rm H}$.
Orbital repulsion is one of the key processes that realize the
 oligarchic growth of protoplanets.
All these elementary processes control the basic mode, timescale, and
 spatial structure of planetesimal accretion. 

On the early stage of planetesimal accretion the growth mode of
 planetesimals is runaway growth where larger planetesimals grow faster
 than smaller ones.
This is due to gravitational focusing that enhances the collisional
 cross-section of planetesimals by the self-gravity. 
The mass distribution of planetesimals is well approximated by a
 power-law distribution of 
 $n \mathrm{d}m \propto m^{-\alpha} \mathrm{d}m$, where 
 $\alpha \simeq 2.5$ for small bodies, and detached runaway bodies.
Once the mass of a protoplanet (runaway-growing body) exceeds a critical
 mass, it effectively heats up the random velocity of local
 planetesimals, which results in orderly growth among protoplanets.
Then orbital repulsion between adjacent protoplanets keeps their orbital
 separations $b \simeq 10 r_\mathrm{H}$.
This growth mode is called as oligarchic growth of protoplanets where
 similar-sized protoplanets predominantly grow with certain orbital
 separations. 
These processes of planetesimal dynamics and accretion are relatively
 well understood and are now incorporated into the standard model of
 planet formation.  

The final stage of terrestrial planet formation is known as giant impact
 stage. 
A protoplanet system becomes orbitally unstable after dispersal of
 disk gas on the timescale 
 $\log t_\mathrm{inst} \simeq c_1 b_\mathrm{ini}/r_\mathrm{H} + c_2$
 where $b_\mathrm{ini}$ is the initial orbital separation of
 protoplanets and $c_1$ and $c_2$ are functions of the initial
 $\langle e^2 \rangle^{1/2}$ and $\langle i^2 \rangle^{1/2}$ and then
 the giant impacts of protoplanets start. 
On this stage accretion proceeds globally and the total mass of
 protoplanets $M_\mathrm{tot}$ is a key parameter that determines the
 mass of planets.  
The masses of the largest and the second-largest planets increase with 
 $M_{\rm tot}$ almost linearly.      
The spin parameters of planets are determined by giant impacts. 
The RMS spin angular velocity $\omega$ is as large as
 70\% of the critical spin angular velocity for rotational instability. 
The obliquity of the planets follow an isotropic distribution.

We also discussed the orbital evolution of planets by the planet-gas
 disk interaction. 
The core accretion model for gas giant formation is summarized and used
 to discuss the diversity of planetary systems.

As we have shown, planet formation consists of multi-scale, multi-layer
 processes regulated by a variety of physics.
Some of fundamental processes are still unclear, in particular,
 formation of planetesimals from dust and orbital migration of
 (proto)planets. 
Both processes are regulated by structure and evolution of
 a protoplanetary disk.
We hope ALMA will reveal detailed structure/evolution of protoplanetary
 disks. 
The diversity of observed extrasolar planets suggests that
 secular gravitational perturbations among planets also
 play an important role in creating the architecture of planetary
 systems. 
We need to deeply explore secular orbital dynamics as well as
 planet-disk interactions to understand the diversity of planetary
 systems. 
We can calibrate theoretical models by using the data of rapidly
 developing observations of extrasolar planets.  
In the theoretical model of planet formation, there are many unsolved
 problems related with many different kinds of physics. 
However, we already have tools to attack the problems.
Studies on planet formation is now on an exciting, rapidly developing
 stage. 

%

%

%

\end{document}